\newcommand\ignore[1]{}

\documentclass[a4paper,11pt]{article}
\pdfoutput=1
\usepackage{graphicx}
\usepackage{jheppub}
\usepackage{color}
\usepackage{amssymb}
\usepackage{amsmath}
\usepackage{amssymb}
\usepackage{tocloft}
\usepackage{float}
\usepackage{braket}
\usepackage{url}
\usepackage{subcaption}
\usepackage{upgreek}
\usepackage{outlines}
\usepackage{braket}
\usepackage{appendix}
\usepackage{placeins}
\usepackage{multirow}

\def\0{{(0)}}
\def\1{{(1)}}

\def\wf{\ensuremath{\text{Weyl}^4}}

\usepackage{bm}
\usepackage[bottom]{footmisc}

\def\L{\Lambda}
\def\calL{\mathcal{L}}

\def\d{\partial}

\def\mn{{\mu\nu}}

\def\a{\alpha}
\def\b{\beta}
\def\ab{{\alpha\beta}}

\def\eps{{\varepsilon}}
\def\calO{\mathcal{O}}
\usepackage{bm}

\def\m{\mu}
\def\n{\nu}

\def\g{\gamma}
\def\k{\kappa}
\def\piG{\pi G}

\def\t{\tau}

\def\Om{\Omega}

\def\l{\lambda}
\def\th{\theta}

\newcommand{\diagthree}[3]{\ccc #1&&\\&#2&\\&&#3 \cccend}

\def\s{\sigma}

\def\onec{\left(\begin{array}{c}}
\def\cend{\end{array}\right)}
\def\cc{\left(\begin{array}{cc}}
\def\ccend{\end{array}\right)}
\def\ccc{\left(\begin{array}{ccc}}

\def\cccend{\end{array}\right)}
\def\cccc{\left(\begin{array}{cccc}}
\def\ccccc{\left(\begin{array}{ccccc}}

\def\calG{\mathcal{G}}
\def\rads{r_{AdS}}
\def\ccccend{\end{array}\right)}
\def\r{\rho}
\def\t{\tau}

\newcommand{\calN}{\mathcal{N}}

\def\mnrs{{\mu\nu\rho\sigma}}

\usepackage{tikz}
\usepackage{pgfplots}
\usetikzlibrary{matrix,arrows,decorations.pathmorphing}
\usetikzlibrary{patterns}
\usetikzlibrary{shapes.misc}
\usetikzlibrary{trees}
\usetikzlibrary{decorations.pathmorphing}
\usetikzlibrary{shapes.geometric}
\usetikzlibrary{positioning}
\usetikzlibrary{decorations.markings}
\usetikzlibrary{positioning,arrows}
\newcommand\numberthis{\addtocounter{equation}{1}\tag{\theequation}}
\usetikzlibrary{decorations.pathreplacing}
\usetikzlibrary{decorations.pathmorphing}
\usetikzlibrary{shapes}
\usetikzlibrary{plotmarks}
\usetikzlibrary{decorations.markings}
\tikzset{
particle/.style={thin,draw=black, postaction={decorate},
decoration={markings,mark=at position .5 with {\arrow[black, line width=0.5mm]{stealth}}}},
gluon/.style={decorate, draw=black, decoration={coil,amplitude=4pt, segment length=5pt}},
photon/.style={decorate, decoration={snake}},
singularity/.style={decorate, draw=black, decoration=zigzag}
}
\tikzstyle{pomeron} = [thin,draw=black]
\tikzstyle{anti} = [thin,draw=black, postaction={decorate},
decoration={markings,mark=at position .5 with {\arrow[black, line width=0.5mm]{stealth reversed}}}]
\tikzstyle{antivirasoro} = [thin,draw=black, postaction={decorate},
decoration={markings,mark=at position .333 with {\arrow[black, line width=0.5mm]{stealth reversed}}}]
\tikzstyle{antivirasoro2} = [thin,draw=black, postaction={decorate},
decoration={markings,mark=at position .666 with {\arrow[black, line width=0.5mm]{stealth reversed}}}]
\tikzstyle{virasoroparticle} = [thin,draw=black, postaction={decorate},
decoration={markings,mark=at position .333 with {\arrow[black, line width=0.5mm]{stealth}}}]
\tikzstyle{virasoroparticle2} = [thin,draw=black, postaction={decorate},
decoration={markings,mark=at position .666 with {\arrow[black, line width=0.5mm]{stealth}}}]

\newcommand{\eq}[1]{\begin{align}#1\end{align}}
\newcommand{\seq}[1]{\begin{align*}#1\end{align*}}
\newcommand{\subeqs}[1]{\begin{subequations}\begin{align}#1\end{align}\end{subequations}}

\def\de{\delta}

\usepackage[bottom]{footmisc}

\title{Stringy Effects and the Role of the Singularity in Holographic Complexity}

\author[a]{Richard Nally}
\affiliation[a]{Department of Physics, Stanford University}
\emailAdd{rnally@stanford.edu}

\numberwithin{equation}{section}
\def\cdot{\dot{C}}
\def\vdot{\dot{V}}
\def\calV{\mathcal{V}}
\def\prs{\left(r_*\right)}
\def\rh{{r_H}}
\def\wf{Weyl$^4$ }
\def\cvsp{\left(\calV^2\right)'}

\begin{document}


\abstract{There has been considerable recent interest in holographic complexity.  The two leading conjectures on this subject hold that the quantum complexity of the boundary thermofield double state should be dual to either the volume of the Einstein-Rosen bridge connecting the two sides (CV conjecture) or to the action of the Wheeler-de-Witt patch of the bulk spacetime (CA conjecture). Although these conjectures are frequently studied in the context of pure Einstein gravity, from the perspective of string theory it is also natural to consider models of gravity in which general relativity is perturbed by higher powers of the Riemann tensor, suppressed by powers of the string length; in a holographic context, these corrections are dual to corrections in inverse powers of the 't Hooft coupling. In this paper, we investigate the CV and CA conjectures in two stringy models of higher-curvature gravity. We find that the CV complexification rate remains well-behaved, but conversely that these corrections induce new divergences in the CA complexification rate that are absent in pure Einstein gravity. These divergences are intrinsically linked to the singularity, and appear to be generic in higher curvature theories. To the best of our knowledge, infinities originating at the singularity have not yet been observed elsewhere in the literature. We argue that these divergences imply that, in the CA picture, the complexification rate of the boundary theory is a nonanalytic function of the 't Hooft coupling.}

\maketitle

\section{Introduction}
The study of quantum gravity has recently been reinvigorated by the introduction of a variety of ideas from quantum information theory, amongst the most exciting of which is the notion of quantum complexity \cite{Susskind:2014rva}. Complexity is an extremely natural quantity in quantum computation, and intuitively describes how difficult it is to prepare some quantum state \cite{Nielsen:2011:QCQ:1972505}. More precisely, fix some set $\{g_i\}$ of ``simple" operators, known as gates, on a Hilbert space $\mathcal{H}$, and two states $\ket{\psi}$ and $\ket{\phi}$. The ``relative complexity" of $\ket{\psi}$ and $\ket{\phi}$ is defined as the smallest number of gates by which we can go from $\ket{\psi}$ to a state sufficiently close to $\ket{\phi}$, i.e. within a small distance $\eps$ of $\ket{\phi}$. We frequently will simply speak directly of the complexity of $\ket{\psi}$, which is simply its complexity relative to some fixed reference state $\ket{\psi_0}$.  

For purposes of holography, it is most interesting to consider the complexity of a state $\ket{\psi(t)}$ as it undergoes unitary time evolution. For a general quantum system with continuous time, this problem is too difficult to be tractable. However, for a particularly simple class of systems known as random quantum circuits, we can study the time evolution of complexity, also known as complexification, rather explicitly; see e.g. \cite{Aaronson:2016vto,Brown:2017jil,Susskind:2018pmk} for reviews\footnote{This approach is known as ``circuit complexity," and although conceptually simple does not generalize cleanly to systems with continuous time evolution. In that more general context, a more promising approach is given by Nielsen geometry \cite{2005quant.ph..2070N,2007quant.ph..1004D}, which defines complexity in terms of geodesic motion on the space of unitary operators. However, this additional complication is unnecessary for our purposes, and we will work only with circuit complexity here. For applications of Nielsen geometry to holography complexity, see e.g. \cite{Susskind:2014jwa,Brown:2016wib}.}. Consider a set of $N$ qubits with state $\ket{\psi(t)}$, with time evolution occuring in discrete chunks. At each time step, we model time evolution by acting on the qubits with $n<N$ randomly chosen gates, each of which only acts on two qubits at a time. We therefore have \eq{\ket{\psi(t)} = g_ng_{n-1}\cdots g_1\ket{\psi(t-1)}.} We are interested in computing the complexity $C(t)$ of $\ket{\psi(t)}$ relative to the initial state $\ket{\psi(0)}$, which we take to be ``generic" in the Hilbert space. For large $N$, it turns out that the details of $C(t)$ are essentially independent of both $\ket{\psi(0)}$ and on the set $\{g_i\}$ of gates. 

\begin{figure}
\begin{center}
\includegraphics{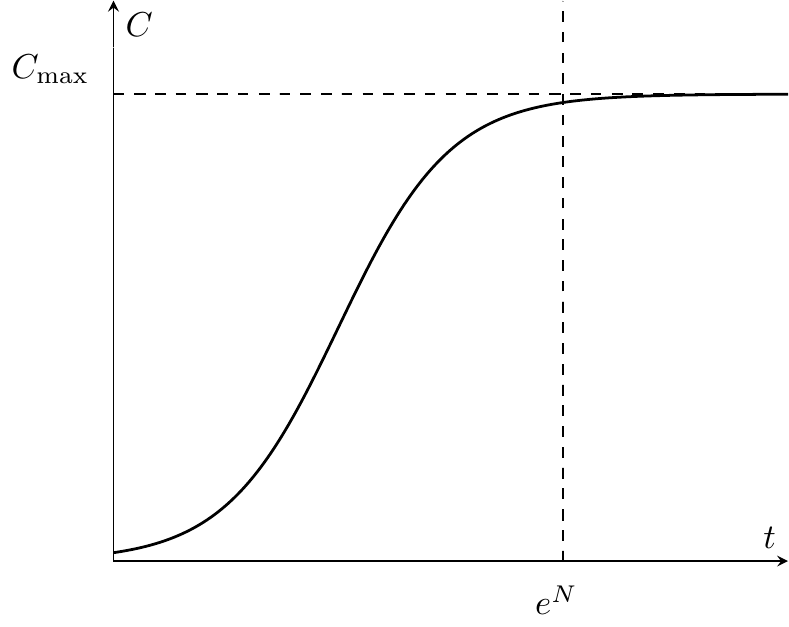}
\caption{A sketch of the time evolution of the complexity $C(t)$ of a random quantum circuit.}
\label{fig:c(t)}
\end{center}
\end{figure}

The complexity of any state relative to itself is zero, so we have $C(t=0)=0$. For generic circuits, after a short period of transient initial behavior, the complexity $C(t)$ grows linearly in time until, at times exponential in $N$, it saturates at a maximum $C_{\text{max}}$. It stays near $C_{max}$ until times doubly exponential in $N$, at which time $C(t)$ begins to decrease linearly with time. This continues until $C(t)$ again reaches 0, and then the entire process begins anew. The behavior of $C(t)$ until the saturation time is sketched in Figure \ref{fig:c(t)}. Importantly, both the saturation of $C(t)$ at $C_{\text{max}}$ and its recurrences in double exponential time are consequences of having a finite dimensional Hilbert space, and thus will not appear in the limit of strictly infinite $N$.

Because in a holographic context we will always take $N$ to be very large, we will primarily be interested in the period of linear growth. There is a simple heuristic to determine the rate of complexification during this regime: the rate is simply the number of qubits performing computation times the average energy per qubit. More simply, we have \cite{Susskind:2014rva} \eq{\cdot \sim ST, \label{eq:dCdt=ST}} where $S$ is the entropy of the system and $T$ is its temperature. It was argued in \cite{2000Natur.406.1047L} that the complexification rate of a system of energy $E$ is always bounded from above by \eq{\cdot \le \frac{2E}{\pi}.\label{eq:lloyd}} This ``Lloyd bound" is conjectured to apply to any quantum system. However, more recent developments have cast the exact prefactor of this bound into question (see e.g. \cite{Cottrell:2017ayj,Carmi:2017jqz,Kim:2017qrq}), and we will treat it simply as the statement that quantum complexity should always develop at a finite rate. 

\begin{figure}
\begin{center}
\includegraphics{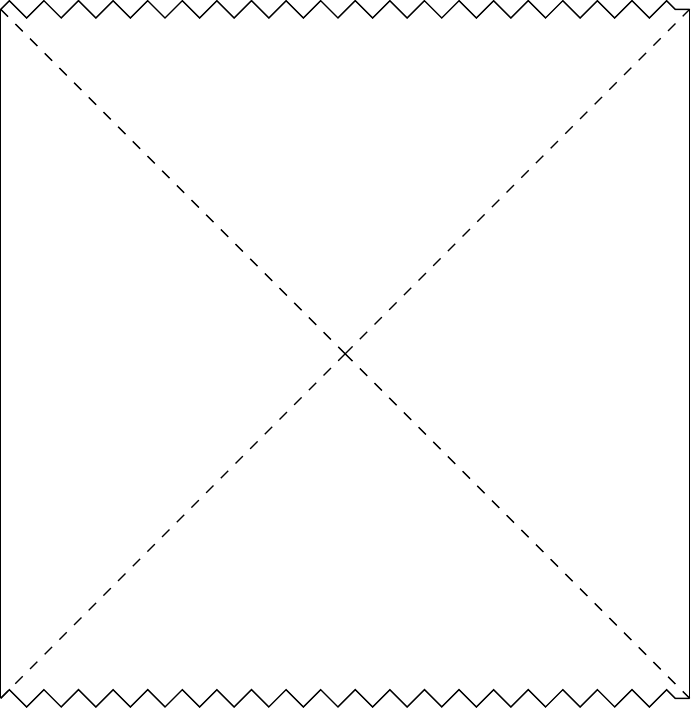}
\caption{The Penrose diagram of an eternal AdS-Schwarzschild black hole.}
\label{fig:penrose}
\end{center}
\end{figure}

Complexity in the context of AdS/CFT was first discussed in \cite{Susskind:2014rva}. As is well-known, AdS/CFT relates the dynamics of a $(d-1)$-dimensional conformal field theory (CFT) to the dynamics of type IIB string theory in $d$-dimensional anti-de-Sitter (AdS) space \cite{Maldacena:1997re,Witten:1998qj,Gubser:1998bc}. In this vein, the goal of holographic complexity is to relate the complexity $C(t)$ of the boundary state to some aspect of the geometry of its bulk dual. For systems as complicated as CFTs, especially in the strong coupling regime where holographic behavior occurs, there is not yet a suitable definition of complexity, so we will frequently take Eq. \ref{eq:dCdt=ST} as the \textit{definition} of complexity. We will restrict ourselves to studying the thermofield double state of the boundary CFT, defined as \eq{\ket{\text{TFD}} = \frac{1}{\sqrt{Z(\b)}} \sum_n e^{-\b E_n} \ket{n}_L\ket{\bar{n}}_R,} where $n$ runs over the energy eigenstates of the boundary theory. For this state, the bulk geometry takes the form of a two-sided eternal AdS-Schwarzschild black hole \cite{Maldacena:2001kr}. In $d$ dimensions, the AdS-Schwarzschild metric is given by \eq{ds^2 = -f(r)dt^2+\frac{dr^2}{f(r)}+r^2d\Om_{d-2}^2,} where the $d$-dimensional emblackening factor is given by \eq{f(r) = 1 - \frac{16\pi M}{(d-2)\Om_{d-2}r^{d-3}} + \frac{r^2}{\rads^2},} where $M$ is the mass of the black hole, set by the temperature $\b$ of the boundary theory, and $\rads$ is the AdS radius, given in terms of the cosmological constant $\L$ by \eq{\L =  -\frac{(d-1)(d-2)}{2\rads^2}.} This geometry has a single horizon at $r=r_H$, defined by the condition \eq{f\left(r_H\right)=0,} and a singularity at $r=0$; its Penrose diagram is shown in Figure \ref{fig:penrose}. There are two leading proposals for how to relate the complexity of a boundary state to the geometry of its bulk: the complexity-volume (CV) conjecture \cite{Susskind:2014rva,Stanford:2014jda,Susskind:2014moa} and the complexity-action (CA) conjecture \cite{Brown:2015bva,Brown:2015lvg}. We will spend the remainder of the paper discussing these two conjectures, so before we move on we will briefly introduce them; a more detailed review is given in Appendix \ref{sec:review}, where we explicitly work out the calculations associated to these two frameworks in the case of pure Einstein gravity.

\begin{figure}
\begin{center}
\includegraphics{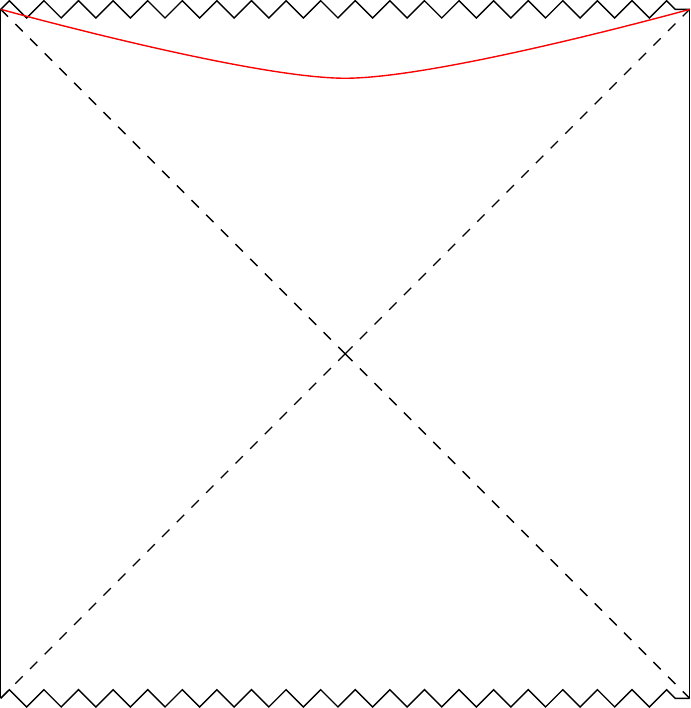}
\caption{The basic setup of the CV conjecture. At very late times, the complexification rate $\cdot_V$ is proportional to the volume of the late-time maximal hypersurface at $r=r_*$, drawn in red. }
\label{fig:CV}
\end{center}
\end{figure}

The first conjecture we will consider is the complexity-volume (CV) conjecture, first introduced in \cite{Susskind:2014rva}. According to the CV conjecture, the complexity $C$ of the boundary state is dual to the volume $V$ of the Einstein-Rosen bridge (ERB) in the TFD geometry of the bulk dual, as shown in Figure \ref{fig:CV}. The volume $V(t)$ of the ERB is defined as the volume of a maximal spatial hypersurface at time $t$. More precisely, we have \eq{C_V = \frac{V}{G\rads}} so that \eq{\dot{C}_V = \frac{\dot{V}}{G\rads}.} At late times, the ERB is approximately a constant-$r$ hypersurface, whose position $r(t)$ varies with time. In this way, the time evolution of the location and size of the ERB causes the time evolution of complexity. Once the late-time behavior takes over, we have that \eq{\dot{V}(t) = \Om_3r^3(t)\sqrt{|f\left[r(t)\right]}} for any fixed time $t$. At very late times, $r(t)$ saturates to a fixed position $r_*$, which is defined to extremize the ``volume functional" $\calV(r)$ given by \eq{\calV(r) = r^3\sqrt{|f(r)|}.\label{eq:calV}} Thus, for late times the volume growth rate is given by \eq{\dot{V} = \Om_3r_*^3\sqrt{|f\left(r_*\right)|} \equiv \Om_3\calV\left(r_*\right),\label{eq:VdotDef}} so that we have \eq{\cdot_V = \frac{\Om_3}{G\rads}\calV\left(r_*\right).\label{eq:CV}}

\begin{figure}
\begin{center}
\includegraphics{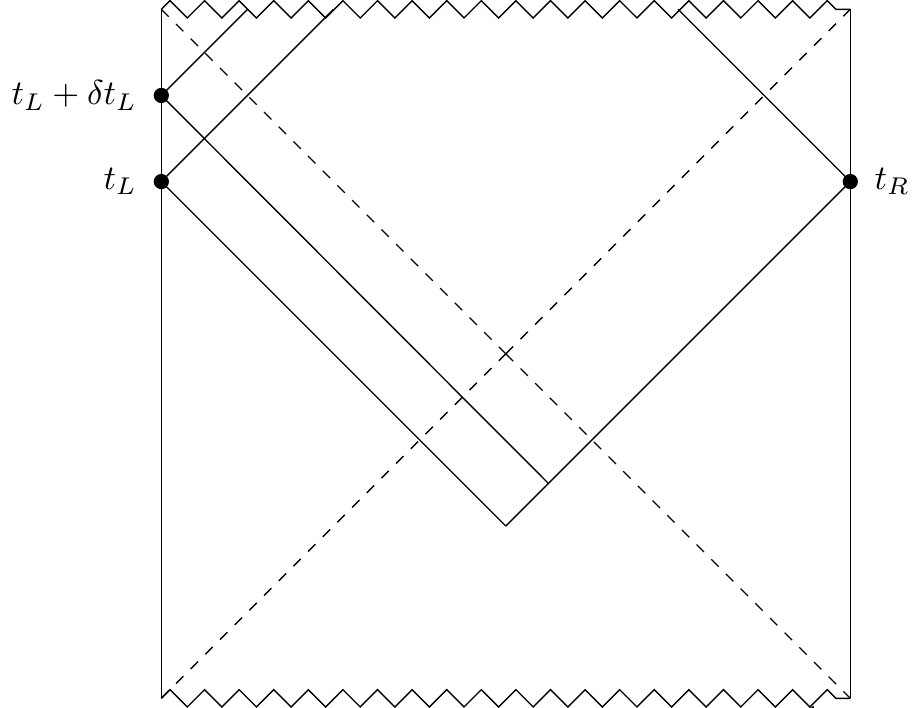}
\caption{Two WdW patches in an AdS-Schwarzschild spacetime. As we evolve the boundary time $t_L$, the WdW patch moves upwards, which determines the time dependence of complexity.}
\label{fig:WdW}
\end{center}
\end{figure}

The other framework for holographic complexity that we will discuss is the complexity = action (CA) conjecture, first introduced in \cite{Brown:2015bva}. The CA conjecture holds that the complexity $C(t)$ of the boundary state is proportional to the action $S_{\text{WdW}}$ of the ``Wheeler-de Witt (WdW) patch," defined as the intersection of the forwards and backwards light cones of two boundary times $t_L$ and $t_R$; we have sketched  WdW patches anchored at successive boundary times $t_L$ and $t_L + \de{t}$ in Figure \ref{fig:WdW}. From the figure, we can see how complexity evolves in time in this context: as we evolve the boundary time forwards, the WdW patch moves, and therefore its associated action changes. We therefore have \eq{\cdot_A = \frac{1}{\pi}\frac{d S_{\text{WdW}}}{dt_L}.\label{eq:CA}} The factor of $\pi$ is chosen such to match the prefactor in Eq. \ref{eq:lloyd}. 

\begin{figure}
\begin{center}
\includegraphics{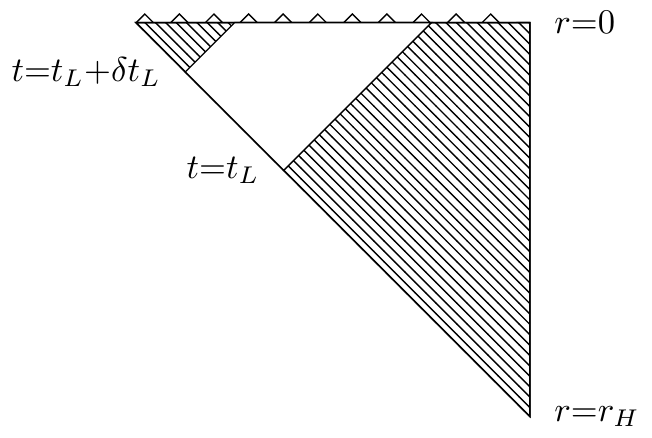}
\caption{The $\de$WdW region, shown unshaded. At late times, it was argued in \cite{Brown:2015bva,Brown:2015lvg} and verified in \cite{Lehner:2016vdi,Carmi:2016wjl} that this region controls $\cdot_V$.}
\label{fig:deWdW}
\end{center}
\end{figure}

In principle, evaluating $\cdot_A$ is quite involved. In general, in addition to the familiar bulk action of general relativity, \eq{S_{\text{GR, bulk}} = \frac{1}{16\piG}\int d^dx\sqrt{-g}\left[R-2\L\right],\label{eq:EH}} we have null boundaries, corner terms, and other complications. These facets add up to give a highly nontrivial relativity problem. A simpler calculation was suggested in \cite{Brown:2015bva,Brown:2015lvg}, where it was argued that, at late times, only the unshaded region in Figure \ref{fig:deWdW} contributes to the time derivative; for the remainder of the paper, we will refer to this region as the  $\de$WdW region, and correspondingly denote its action by $S_{\de\text{WdW}}$. In \cite{Lehner:2016vdi,Carmi:2016wjl}, the full calculation, including a careful treatment of the subtle issue of null boundaries, was performed, and was shown to match the much simpler calculation involving the $\de$WdW patch. We will therefore focus on the $\de$WdW region throughout the paper. From Figure \ref{fig:deWdW}, we can straightforwardly read off that\footnote{It is argued in \cite{Brown:2015bva,Brown:2015lvg} that all corner terms in the $\de$WdW region cancel by time translation invariance, leaving us with only these contributions.} \eq{S_{\de\text{WdW}} = S_{\text{bulk}} + S_{\text{bdy}}\left[r=r_H\right] - S_{\text{bdy}}\left[r=\eps\right],\label{eq:SWdW}} where we have instituted a UV cutoff at $r=\eps$ to avoid the awkward situation of evaluating a boundary term at the singularity\footnote{For more on the role of a cutoff in CA, see e.g. \cite{Akhavan:2018wla,Alishahiha:2019cib,Hashemi:2019xeq}. Such cutoffs are quite relevant to the study of holographic complexity in Jackiw-Teitelboim gravity \cite{Brown:2018bms,Goto:2018iay,Alishahiha:2018swh}.}. We will use this UV cutoff throughout the paper.

The CV and CA conjectures were originally introduced in the context of pure Einstein gravity. However, string theory naturally comes equipped with an infinite tower of corrections to GR, known as $\a'$ corrections (see e.g. \cite{Zwiebach:1985uq,Zumino:1985dp,Gross:1986iv} for historical references, and \cite{Tong:2009np} for a more modern overview). These corrections are suppressed by powers of the string length squared, and take the form of higher curvature theories. In particular, we supplement the Einstein-Hilbert action in Eq. \ref{eq:EH} with a series of corrections of the form \eq{S = \frac{1}{16\piG} \int d^dx\sqrt{-g}\left[R - 2\L + \sum_{n=1}^\infty \a'^n \calL_n\right],} where $\calL_n$ is in general an $n$-th order polynomial in the Riemann tensor and possibly the gauge fields. For the TFD geometry, which corresponds to a neutral black hole, the gauge fields will not contribute, and we can think of the $\a'$ corrections as being built simply out of the Riemann tensor. These corrections come from loop diagrams on the string worldsheet, and are omnipresent in modern string theory. For instance, $\a'$ corrections are absolutely crucial to the modern understanding of the entropies of BPS black holes, which ranks amongst the foremost successes of string theory; see e.g. \cite{SenNotes,DabNotes} for reviews. 

At this point it is useful to emphasize one crucial difference between the CV and CA conjectures. In the CV approach, the essential physics, although occurring inside the horizon, remains well separated from the singularity, as can be seen from Figure \ref{fig:CV}. For large black holes, the critical radius $r_*$ remains near the horizon, and so in a low-curvature region where the gravity approximation can still be trusted. For CA, on the other hand, the singularity is directly involved in the physics. The WdW patch, at least for a neutral black hole, reaches down to the singularity, as can be easily seen in Figures \ref{fig:WdW} and \ref{fig:deWdW}. This is somewhat concerning: as we approach the singularity, the curvature of spacetime becomes strong, and we expect gravitational effective field theory to break down. Nevertheless, it was observed in \cite{Brown:2015bva,Brown:2015lvg} that the contributions to the pure GR action from the singularity are well-behaved, despite naive expectations to the contrary; we review these findings in Appendix A. Once higher curvature effects are taken into consideration, it is unclear whether this well-behavedness at the singularity should still be expected to hold. Indeed, before this work, the question of singularity effects in CA for higher-curvature gravity was fairly unexplored, as most previous work centered on charged black holes, where the WdW patch ends at the inner horizon and so avoids the singularity.

Our goal in this paper is to study the effect of $\a'$ corrections on bulk complexification, in both the CV and CA pictures. Throughout the paper, we will carefully work perturbatively, treating $\a'$ as a small parameter and only working to the first subleading order. This is the correct framework to use from the perspective of string theory, where we always take the string length to be small. On physical grounds, we expect $\a'$ corrections to lower the complexification rate relative to the results in pure GR. We present two arguments justifying this expectation. The first originates in the related area of quantum chaos. It is intuitively clear that notions of quantum chaos in gravity, explored in \cite{Shenker:2013pqa,Shenker:2014cwa,Maldacena:2015waa}, are related to questions of complexity; see e.g. \cite{Susskind:2018tei,Roberts:2018mnp,Brown:2018kvn,Qi:2018bje,Parker:2018yvk} for progress towards making this intuition precise. Chaos can be measured by considering thermal out-of-time-order correlations of the form $\braket{VW(t)VW(t)}_\beta$, where $\beta$ is the inverse temperature. These correlators have a characteristic behavior of the form \eq{\braket{VW(t)VW(t)}_\beta \sim e^{\l t},\label{eq:chaos}} where $\l$ is the called the ``Lyapunov exponent" of the theory. The Lyapunov exponent is bounded from above by \cite{Maldacena:2015waa} \eq{\l \le \frac{2\pi}{\b}.} Classical gravity saturates this bound. The effects of $\a'$ corrections to this correlator were worked out in \cite{Shenker:2014cwa}. It was shown that $\a'$ corrections preserve the exponential decay in Eq. \ref{eq:chaos}, but with the Lyapunov exponent picking up a negative $\a'$ correction given by \cite{Shenker:2014cwa,Maldacena:2015waa} \eq{\l = \frac{2\pi}{\b}\left(1-c\frac{\a'}{\rads^2}\right),} where $c$ is a constant Intuitively, the Lyapunov exponent should play a similar role to the rate of complexification, so we expect similar corrections to appear here. 

For an alternative heuristic, we appeal to a more intuitive picture\footnote{We thank L. Susskind for suggesting this picture to us.}. The reasoning in \cite{Stanford:2014jda,Susskind:2014moa,Brown:2015bva,Brown:2015lvg} and related works that led to the origins of the CV and CA conjectures were based on a Rindler space picture, in which the black hole horizon is infinitely large. Because we are explicitly interested in corrections coming from a finite string length, we cannot take this limit, and instead must consider strictly finite size black holes. To obtain an intuition for what to expect away from this limit, it is instructive to consider the opposite limit, in which the strings are larger than the black hole. In this limit, the probe string wraps around the black hole completely, and so each differential length of the string is attracted to the black hole. Adding up all these attractive forces, we see that the string feels no force, and is essentially free. We expect that free theories should complexify less quickly than strongly interacting theories, so this suggests that stringy effects should lower the complexification rate.

Higher-curvature corrections to bulk complexification have been studied in previous work; see e.g. \cite{Wang:2017uiw,Guo:2017rul,Alishahiha:2017hwg,Cano:2018aqi,Cai:2016xho,An:2018dbz,Cano:2018ckq,Jiang:2018sqj,Jiang:2018pfk,HUANG2017357,Ding:2018ibq,Meng:2018vtl,Bueno2018,Chapman:2016hwi,Chapman:2018dem,Chapman:2018lsv,Cai:2017sjv,Sebastiani:2017rxr,HosseiniMansoori:2018gdu} for a variety of bulk Lagrangians and geometries. We will discuss two higher-curvature theories of gravity in this paper, both of which are directly motivated by string theory. The first is the well-known theory of Gauss-Bonnet (GB) gravity, where we supplement the Einstein-Hilbert Lagrangian with a term appropriate for computing the Euler character of a four-manifold: \eq{\calL \supset \a'\left(R_\mnrs R^\mnrs - 4R_\mn R^\mn + R^2\right).} This is in some sense the canonical higher-curvature theory, as any quadratic polynomial in the Riemann tensor can be repackaged into this form by an appropriate change of variables. The GB Lagrangian also appears naturally in many string compactifications, see e.g. \cite{Sen:2005iz}. Additionally, as an example of a Lovelock theory, the GB theory is an example of a higher-curvature Lagrangian that is \textit{not} a higher-derivative theory. This means in particular that there are no ghosts, and it is possible to render the variational problem for the metric well-posed by adding in an appropriate boundary term \cite{PhysRevD.36.392}. We must mention here that both the CV and CA proposals have already been studied in this theory; see e.g. \cite{An:2018dbz,Cai:2016xho,Cano:2018ckq,Cano:2018aqi}. We will nonetheless work this theory out explicitly, for two reasons. First, our perturbative framework will lead to us reaching qualitatively different conclusions than those references, which worked nonperturbatively $\a'$, and secondly, the GB theory is an excellent warmup for the rather more complicated theory we introduce below.

The GB theory has one drawback for use in holographic contexts: it doesn't correspond in a precise way to a known boundary theory. Even if there were a satisfactory definition of complexity for boundary theories, we wouldn't have a theory with which to compare to bulk results. To remedy this, the second theory we consider will be the known higher-curvature theory which corresponds to the leading finite-'t Hooft correction to four-dimensional $\calN=4$ supersymmetric Yang-Mills (SYM) theory. The 't Hooft coupling $\l$ is defined as \eq{\l \equiv g_{\text{YM}}^2N.} The standard holographic dictionary relates $\l$ to $\a'$ by \eq{\l = \frac{\rads^4}{(\a')^2}.} For the specific case of $\calN=4$ SYM, It is well-known that to compute the leading correction in $\l$, we must perturb the Einstein-Hilbert by a quartic polynomial in the Weyl tensor\footnote{ More precisely, it was shown in \cite{Myers:2008yi,Buchel:2008ae} that, in addition to the expected $\l^{-3/2}$ corrections, this term also is also sufficient to compute corrections on $\sqrt{\l}/N^2$; we will neglect this subtlety in the remainder of this paper. }: \eq{\calL \supset \a'^3\ \text{Weyl}^4.} This theory, which first appeared in \cite{Banks:1998nr}, has been used to compute corrections to the thermodynamics \cite{Gubser:1998nz,Pawelczyk:1998pb}, entanglement entropies \cite{Galante:2013wta}\footnote{This reference computed the Euclidean action of an AdS black hole solution to this theory. This is conceptually similar to the CA analysis we will perform here, but is logically distinct and involves a different set of technical challenges.}, and hydrodynamics \cite{Buchel:2004di,Myers:2008yi,Buchel:2008ae,Buchel:2008vz} of $\calN=4$ SYM, and has been a fertile ground for precision tests of AdS/CFT. Additionally,  quantum chaos in this model \cite{Grozdanov:2018kkt} has been studied the context of pole skipping \cite{Grozdanov:2017ajz}.

However, from a purely bulk perspective, there are issues with this theory. Unlike GB gravity, this theory is a genuine higher-derivative theory. This means in particular that the question of introducing a boundary term is quite subtle \cite{Simon:1990ic,Dyer:2008hb}. However, a very simple boundary term for $f(\text{Riemann})$ theories has recently been derived in \cite{Jiang:2018sqj,Deruelle:2009zk}. We will use this term to facilitate a CA analysis for the Weyl$^4$ theory. To the best of our knowledge, this is the first time that holographic complexity has been considered for this theory, and therefore the first holographic attempt at treating non-universal contributions to complexification in $\calN=4$ SYM. We consider this to  be a ``slam dunk" example of complexification in higher-curvature gravity, and it is the main focus of our paper.

In the remainder of this paper, we will analyze bulk complexification for these two higher-curvature theories, using both the CV and CA proposals.  The results of these analyses are summarized in Table \ref{tab:summary}. These calculations were performed with the Mathematica packages \textit{ccgrg} \cite{Woszczyna:2016zyo,ccgrg} and \textit{xTensor} \cite{xtensor}.

\begin{table}[h]
\begin{centering}
\begin{tabular}{|c|c|c|}\hline
Theory & Complexity-Volume Result & Complexity-Action Result \\\hline
5d Gauss-Bonnet & $\calV(r)$ finite, $\cdot_V$ given in Eq. \ref{eq:cvGB}. & Divergent, with $\cdot_A\sim \a/\eps^4$. \\\hline
\wf & $\calV(r)$ divergent, $\cdot_V$ given in Eq. \ref{eq:cvW}. & Divergent, with $\cdot_A\sim \g/\eps^{12}$. \\\hline
\end{tabular}
\caption{A summary of the results of this paper. We consider two higher-curvature gravity theories, the five-dimensional Gauss-Bonnet theory defined in Eq. \ref{eq:5dGBbulkAction} and the \wf theory defined in Eq. \ref{eq:WbulkAction}. We analyze each of these theories in both the complexity-volume\cite{Stanford:2014jda,Susskind:2014moa} and complexity-action \cite{Brown:2015bva,Brown:2015lvg} frameworks. We find that, although for CV divergences do appear near the singularity in the \wf theory , it is possible to extract a finite correction to $\cdot_V$ in both cases. For CA, conversely, all results are divergent, and these divergences appear to be unavoidable in any neutral black hole background. We argue below that these divergences are generic, and should appear for general higher-curvature theories.}
\label{tab:summary}
\end{centering}
\end{table}

One of the major themes of this paper is that holographic complexity is extremely sensitive to singularity effects and the breakdown of gravitational EFT. In particular, we will see that higher-curvature effects cause divergences to appear in both CV and CA calculations, and moreover that these infinities render the final CA complexification rate divergent. To the best of our knowledge, this is the first appearance of divergences in the complexification rate directly related to the singularity in the holographic complexity literature\footnote{On the other hand, divergences in the complexification rate originating at the asymptotic boundary of AdS space have been observed before \cite{Moosa:2017yiz}, in the context of both CV and CA. The divergences observed there are related to time-dependent perturbations, and do not seem to be related to the breakdown of bulk EFT. Nonetheless, they are extremely interesting for understanding the relationship between the Lloyd bound and the holographic complexity conjectures.}; we anticipate, however, that similar divergences would be found in essentially any other higher-curvature theory, at least in the neutral black hole background.

The divergences we will encounter are in the CA \textit{complexification rate} $\cdot_A(t)$. These stand in distinction to divergences in the total complexity $C(t)$, which have been previously observed in the literature (see e.g. \cite{Carmi:2016wjl,Reynolds:2016rvl}). In terms of the bulk geometry, these divergences in the total complexity are related to the divergent volume of the AdS spacetime. Moreover, it is fairly clear why, from the boundary side, we should expect the total complexity to be divergent: the Hilbert space of a realistic boundary theory is infinite dimensional, so the relative complexity between any two states should generically also be infinite. 

The interpretation of divergences in the complexification rate, on the other hand, is more subtle. The Lloyd bound suggests that any system, even if it has an infinite-dimensional Hilbert space, should have a finite complexification rate. However, from our holographic calculations, it seems that the leading correction to the CA complexification rate is divergent, in stark contrast to this intuition. From this perspective, therefore, it is difficult to make sense of these CA results. 

The corrections we will compute here, although divergent, always enter with a negative sign, as anticipated above. This suggests that, in the context of the CA proposal, all stringy black holes decomplexify infinitely quickly! We will, however, propose an alternate interpretation. Our results imply that, when the viewed as a function of the 't Hooft coupling $\l$, the CA complexification rate $\cdot_A(\l)$ has a well defined value at $\l=\infty$, but a poorly defined first derivative. This is redolent of the behavior of e.g. $\sqrt{x}$ at $x=0$. We therefore believe that our results constitute strong evidence that, if the CA proposal should prove correct, then the complexification rate must be nonanalytic in the 't Hooft, at least around $\l=\infty$. In this way, even though $d\cdot_A/d\l$ is infinite, $\cdot_A(\l)$ itself can be well-defined everywhere.

The outline of this paper is as follows. First, in Section \ref{sec:volume}, we perform the CV analyses for our two higher curvature theories. As part of these calculations, we must necessarily compute corrections to the emblackening factor. We then proceed to use this corrected metric to perform CA analyses for the two theories. First in Section \ref{sec:GB} we perform the CA analysis for Gauss-Bonnet gravity in four and five dimensions. Although the four-dimensional theory is topological, this short exercise will establish conventions and nomenclature before the five dimensional calculations that follow it. Next, in Section \ref{sec:W4} we will perform the CA analysis for the \wf theory that computes the leading-$\l$ correction in $\calN=4$ SYM. This calculation is our main result. Critical to this discussion will be the $f(\text{Riemann})$ boundary term derived in \cite{Deruelle:2009zk,Jiang:2018sqj}. Finally, we conclude with discussion and interpretation in Section \ref{sec:conc}.

\section{CV at Higher Curvature}
\label{sec:volume}
We will begin with exploring stringy corrections to the CV conjecture. As mentioned above, the CV proposal identifies the complexity of the boundary state with the volume of the Einstein-Rosen bridge (ERB) in the TFD geometry. To evaluate this in the higher curvature gravity theories, we will need to compute the perturbed emblackening factor. In studying the corections to the metric, we will follow the framework laid out in \cite{Galante:2013wta}. Following that paper, we will parameterize the perturbed metric as \eq{ds^2 = -a(r)dt^2 + b(r)dr^2 + r^2d\Om_3^2,\label{eq:abMetric}} where \begin{subequations}\label{eq:abDefEps}\begin{align}a(r) &= f_0(r)\left[1+\eps f_1(r)\right] \\ b(r) &= \frac{1}{f_0(r)\left[1+\eps f_2(r)\right]} \sim \frac{1}{f_0(r)}\left[1-\eps f_2(r)\right].\end{align}\end{subequations} Here, $\eps$ is a dimensionless expansion parameter that depends on the theory in question, and $f_0$ is the unperturbed emblackening factor given in Eq. \ref{eq:5demblackening}. This parameterization is chosen so that the coordinate position of the horizon does not change: $a(r)$ and $b(r)$ vanish where $f_0(r)$ does, i.e. at the event horizon $r=r_H$ defined in Eq. \ref{eq:rH}. In this section, we will first compute the corrected metric and then from the metric compute the corrected extremal radius $r_*$ and complexification rate $\cdot_V$ for our two higher-curvature theories. 

For a metric of the form in Eq. \ref{eq:abMetric}, the ``volume functional" defined in Eq. \ref{eq:calV} takes the form \eq{\calV = r^3\sqrt{|a(r)|}.} To recap briefly, we need to find the extrememum of this functional, which we denote by \eq{r_* = r_0 + \a r_1,\label{eq:rstar}} where $r_0$ is defined in Eq. \ref{eq:r0}. To avoid dealing with the square root, it will actually be more convenient to extremize the square of the volume functional, \eq{\calV^2 = -r^6a(r).} Taking the derivative, we have \eq{\left(\calV^{2}\right)'(r)= -6r^5a(r) - r^6a'(r) = -6r^5f_0(r) - r^6f_0'(r) - \eps\left[6r^5f_0'(r)f_1(r) + r^6f_0'(r)f_1(r) + r^6f_0(r)f_1'(r)\right].} We therefore have that {\scriptsize \eq{\cvsp\left(r_*\right) &= -6r_0^5f_0\left(r_0\right) - r_0^6f_0'\left(r_0\right)-\eps\left[r_0^4 \left(r_0 \left(r_0 r_1 f_0''(r_0)+f_0'(r_0) (r_0 f_1(r_0)+12 r_1)\right)+f_0(r_0) \left(r_0^2 f_1'(r_0)+6 r_0 f_1(r_0)+30 r_1\right)\right)\right] + \calO(\eps^2).}} At the extremum, this derivative should vanish to order $\eps$. The $\calO(\eps^0)$ piece automatically vanishes by the definition of $r_0$. and we can make the $\calO(\eps)$ piece vanish by setting\footnote{If we instead extremize $\calV$ directly, we obtain a slightly different expression for $r_1$. However, to $\calO(\eps)$, the two are equivalent, as can easily be verified by inserting the form of $f_0$ and $f_1$.} \eq{r_1 = -\frac{r_0^2f_1\left(r_0\right)f_0'\left(r_0\right) + r_0^2f_0\left(r_0\right)f_1'\left(r_0\right) + 6r_0f_0\left(r_0\right)f_1\left(r_0\right)}{r_0^2 f_0''(r_0)+12 r_0 f_0'(r_0)+30 f_0(r_0)}.} Plugging back into Eq. \ref{eq:calV}, we then have that \eq{\calV\left(r_*\right) = r_0^3 \sqrt{f_0\left(r_0\right)}+\eps\frac{  r_0^3 r_1 f_0'\left(r_0\right)+r_0^3 f_0\left(r_0\right) f_1\left(r_0\right)+6 r_0^2 r_1 f_0\left(r_0\right)}{2 \sqrt{f_0\left(r_0\right)}}+\calO\left(\alpha ^2\right),} and therefore that \eq{\cdot_V = \frac{\Om_3}{G\rads}\left[r_0^3 \sqrt{f_0\left(r_0\right)}+\eps\frac{  r_0^3 r_1 f_0'\left(r_0\right)+r_0^3 f_0\left(r_0\right) f_1\left(r_0\right)+6 r_0^2 r_1 f_0\left(r_0\right)}{2 \sqrt{f_0\left(r_0\right)}}\right]+\calO\left(\alpha ^2\right).\label{eq:cvCorrectionAbstract}}

In this section, we will carry out this analysis for first the Gauss-Bonnet theory and then the \wf theory, with surprising results. Although for GB we will find that the physics is quite similar to the pure gravity case, for \wf we will see that $\calV(r)$ diverges as we approach the singularity, and therefore that radically different physics emerges. To understand how this happens, it is instructive to consider instead the case of pure general relativity, for which it doesn't happen. For pure GR, we have \eq{\calV(r) = r^3\sqrt{-1+\frac{\m}{r^2}-\frac{r^2}{\rads^2}}.} For small $r$, $\calV(r)$ scales as \eq{\calV(r) \sim r^3\frac{1}{r} \sim r^2,} so $\calV(r)$ vanishes at the singularity. More geometrically, this means that the vanishing of the three-sphere volume at the singularity overwhelms the divergence in $f_0$, keeping the late-time slice shown in Figure \ref{fig:CV} away from the singularity. However, this need not be the case in general. For higher-curvature theories, the perturbed emblackening factors will diverge more strongly at the singularity. For sufficiently large powers of $r$, this divergence can instead overwhelm the vanishing of the three-sphere, thus leading $\calV$ to be divergent at the singularity, and moving $r_*$ to $r=0$. This explains the difference between the two theories: in the GB case we will have $f_1\sim r^{-4}$, but instead we will have $f_1\sim r^{-12}$ for the \wf theory. Based on simple dimensional analysis, we expect the behavior of the \wf theory to be prevalent in generic higher-curvature theories: higher powers of $\a'$ correspond in general to more negative powers of $r$ in $f_1$, and therefore drag us towards the divergent regime.

In principle, there might be corrections to $\cdot_V$ other than those that we have considered here\footnote{We thank R. Myers for enlightening comments on this point.}. To see why, it is convenient to make an analogy to black hole entropy. The Bekenstein-Hawking entropy \eq{S_{BH} = \frac{A}{4G}} is completely universal, but to compute higher-curvature corrections to this entropy we must instead use the Wald entropy \cite{Wald:1993nt}, \eq{S_{\text{Wald}} = 2\pi \int_{\text{horizon}} d^{d-2}x\sqrt{-\g} \frac{d\calL}{dR_\mnrs} \eps^{\m\r}\eps^{\n\s},} where $\g_{ij}$ is the induced metric on the horizon, $\calL$ is the Lagrangian of our higher-curvature theory, and $\eps^{\m\r}$ is an antisymmetric two-tensor with components along the $t$ and $r$ axes. In terms of the area $A$ of the black hole, this entropy admits an expansion of the form  \cite{SenNotes,DabNotes} \eq{S_{\text{Wald}} = \frac{A}{4G} + \a'c_1\log{A} + \a'c_2 + \left(\a'\right)^2\frac{c_3}{A} + \cdots.} Computing the entropy of a black hole amounts to finding values for the proportionality constants $c_i$. One might expect that, in analogy with black hole entropy, there should exist a ``Wald complexity" of the form \eq{\cdot_{V,\text{ Wald}} = \frac{1}{G\rads}\left(1 + \a' c_1\log{V} + \a'c_2 + (\a')^2\frac{c_3}{V} + \cdots\right).} However, in the absence of a Wald complexity we do not know how to compute these corrections in the first place, so of course we cannot evaluate them, and Eq. \ref{eq:cvCorrectionAbstract} is the best that we can currently do. See, however, \cite{Bueno:2016gnv,Belin:2018fxe,Belin:2018bpg}, for progress on finding these corrections.

\subsection{CV for Gauss-Bonnet Gravity}
We will now perform the above analysis for perturbative GB gravity. In order to compute the corrected complexification rate for GB gravity, we must first find the correction to the metric. In four dimensions, the GB term is topological, and thus the metric does not get corrected. From Eq. \ref{eq:CV}, we therefore can easily see that there is no correction to the CV rate. We are therefore led to consider naturally the GB theory in five dimensions. For this theory, the expansion parameter $\eps$ used in Eq. \ref{eq:abDefEps} is defined as \eq{\eps \to \a \equiv \frac{\a'}{\rads^2},\label{eq:aDef}} where we have rescaled $\a'$ to give a dimensionless expansion parameter. In terms of $\a$, the five-dimensional GB action is \cite{GBpaper,PhysRevD.36.392,0264-9381-27-17-175014,An:2018dbz,Cai:2016xho,Cano:2018aqi} \eq{S_{\text{GB, bulk}} = \frac{1}{16\piG}\int d^5x\sqrt{-g}\left[R - 2\L + \a\rads^2\left(R_\mnrs R^\mnrs - 4R_\mn R^\mn + R^2\right)\right],\label{eq:5dGBbulkAction}} where \eq{\L = -\frac{6}{\rads^2}} is the cosmological constant. 

To find the forms of $f_1$ and $f_2$, we will follow the strategy laid out in \cite{Galante:2013wta} and plug the metric ansatz, including the expansions of $a$ and $b$ and the known form of $f_0$, into Eq. \ref{eq:5dGBbulkAction}. The entire action will only depend on $r$, so we will obtain an effective one-dimensional action for $f_1$ and $f_2$, from which we can obtain Euler-Lagrange equations which we can solve for $f_1$ and $f_2$. It is essential that we expand the action to second order in $\a$, as the expansion to first order will vanish since $f_0$ minimizes the EH Lagrangian.

Doing so, we find the EL equations \begin{subequations}\label{eq:ELeqsGB} \begin{align} 0&= -3\frac{f_2'(r) \left[r^9+r^7 \rads^2-\mu  \left(r^5 \rads^2\right)\right]+f_2(r) \left[\left(2 r^6\right) \left(2 r^2+\rads^2\right)\right]-8 r^8+\mu ^2 \left(8 \rads^4\right)}{2 r^5 \rads^2}+O\left(\alpha\right)\\ 0 &= 3\frac{f_1'(r) \left[r^9+r^7 \rads^2-\mu  \left(r^5 \rads^2\right)\right)+f_2(r) \left(\left(2 r^6\right) \left(2 r^2+\rads^2\right)\right)-8 r^8+\mu ^2 \left(8 \rads^4\right)}{2 r^5 \rads^2}+O\left(\alpha\right).\end{align}\end{subequations} These EL equations have solutions \begin{subequations}\label{eq:gbf1f2solnC1C2} \begin{align}f_1(r) &=  \frac{C_1}{-\rads^2 \mu +\rads^2 r^2+r^4}+\frac{2 \left(\rads^4 \mu ^2+r^8\right)}{\rads^2 \left(r^6-\mu  r^4\right)+r^8} + C_2 \\ f_2(r) &= \frac{C_1}{-\rads^2 \mu +\rads^2 r^2+r^4}+\frac{2 \left(\rads^4 \mu ^2+r^8\right)}{\rads^2 \left(r^6-\mu  r^4\right)+r^8},\end{align}\end{subequations} where $C_1$ and $C_2$ are integration constants which we now need to fix. We will follow the conventions for these constants set in \cite{Galante:2013wta}. In particular, $C_2$ is simply a time reparameterization, so we are free to set it to zero; we fix $C_1$ so that $f_1$ and $f_2$ are finite at the horizon. This gives us that \eq{C_1 = -\frac{2\left(\rads^4\m^2+r_H^8\right)}{r_H^4},} where $r_H$ is defined in Eq. \ref{eq:rH}. Putting it all together, we have that \eq{f_1(r) = f_2(r) = \frac{2\left(r_H^4-r^4\right)\left(\rads^4\m^2-r^4r_H^4\right)}{r^4r_H^4\left[r^4+\rads^2\left(r^2-\m\right)\right]}.\label{eq:GBf1f2}}

It is worth mentioning that the form of the metric derived here is an exact solution to  GB gravity. The action in Eq. \ref{eq:5dGBbulkAction} has equations of motion given by \cite{GBpaper} \eq{R_\mn - \frac{1}{2}Rg_\mn + \Lambda g_\mn = 8\pi\a'T_\mn,} where we have defined an effective stress tensor $T_\mn$ by \eq{T_\mn = g_\mn\left(R_{\a\b\gamma\de}R^{\a\b\gamma\de} - 4R_\ab R^\ab + R^2\right) - 4RR_\mn + 8R_{\m\a}{R^\a}_\n + 8R_\ab {{{R^\a}_\m}^\b}_\n - 4R_{\m\a\b\g}{R_\n}^{\a\b\gamma}\label{eq:stresstensor}.} For spacetime dimension $d>4$, a general  AdS-Schwarzschild solution to these equations has been found \cite{GBpaper,0264-9381-27-17-175014}. It is of the usual spherically symmetric and static form given in Eq. \ref{eq:abMetric}, with $a$ and $b$ given by \eq{a(r) = b(r) = 1 + \frac{r^2}{2\tilde{\a}}\left[1\mp \sqrt{1+4\tilde{\a}\left(\frac{\m}{r^{d-1}} - \frac{1}{\rads^2}\right)}\right]\label{eq:gbMetricNonlinear},} where \eq{\tilde{\a} = \a\rads^2(d-3)(d-4).} This is the exact solution to the GB equations of motion, and was studied in a CV context in e.g. \cite{An:2018dbz}. However, this metric is nonlinear in $\a$. Because we are imagining the GB term as the leading term in an infinite series of $\a'$ corrections to Einstein gravity, however, this form of the metric is unsuitable for our purposes, as it consists of an entire power series in $\a'$. Instead, we should linearize this solution, which returns us to the form of the metric derived above.  Therefore, the metric given in Eq. \ref{eq:GBf1f2} is the appropriate form of the metric for studying stringy corrections.  

\begin{figure}
\begin{centering}
\includegraphics[scale=1.4]{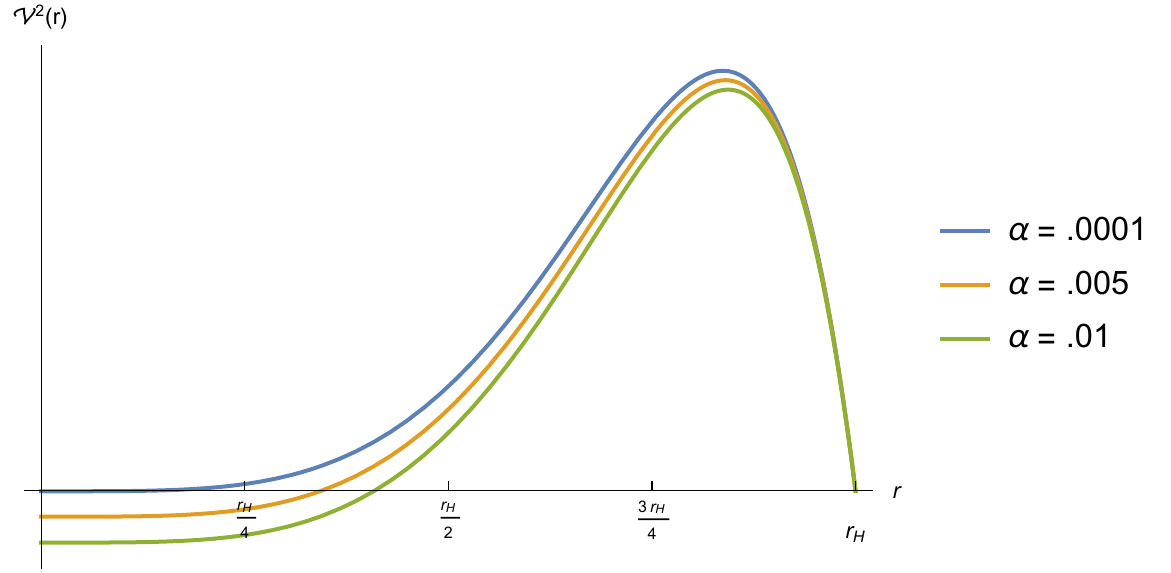}
\caption{The square $\calV^2(r)$ of the volume functional for the $\a$-corrected AdS-Schwarzschild geometry in the GB theory, plotted for several values of $\a$ (labeled).}
\label{fig:cvGB}
\includegraphics[scale=1.4]{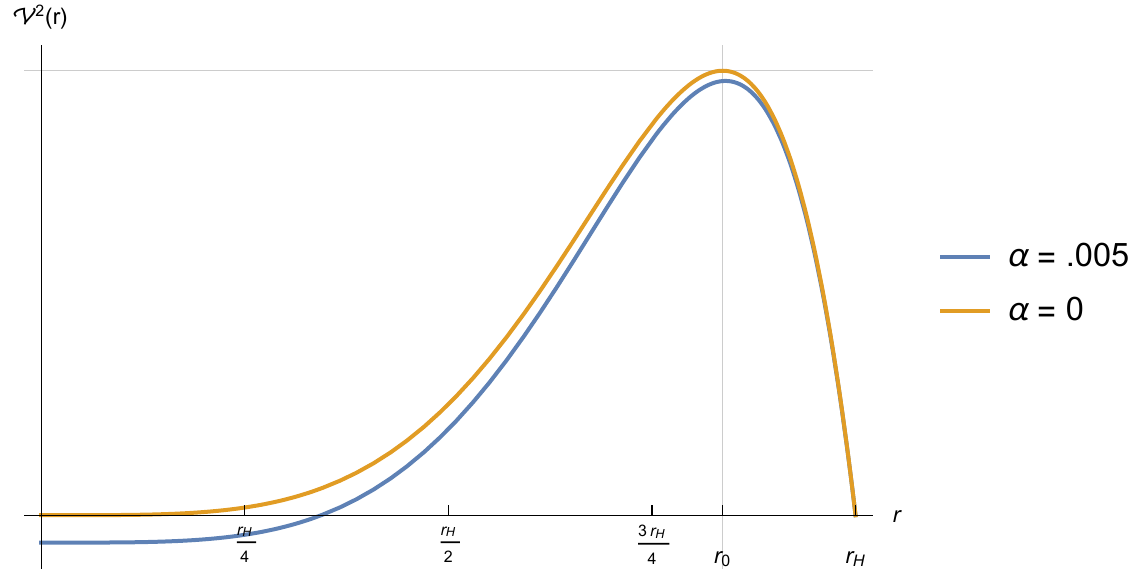}
\caption{The square $\calV^2(r)$ of the volume functional for both the $\a$-corrected AdS-Schwarzschild geometry and the uncorrected geometry. We can see immediately that the $\a'$ correction reduces the complexification rate, as expected.}
\label{fig:cvGBGR}
\end{centering}
\end{figure}

Using the metric in Eq. \ref{eq:GBf1f2}, we can straightforwardly compute $\calV(r)$, which is given by \eq{\calV(r) = \sqrt{-f_0(r)}\left(r^3 + \a\frac{r^4 - r^2\rads^2-\rads^2\m}{r}\right) + \calO(\a)^2.} The square $\calV^2(r)$ of the volume functional is plotted in Figure \ref{fig:cvGB}\footnote{This plot uses the numerical values $\rads=10,000$ and $\m=1,000,000,000$. We will use these values throughout the paper.} for several values of $\a$.

 We find similar physics to the familiar CV analysis of pure GR, as can be seen directly in Figure \ref{fig:cvGBGR}. We see a single maximum inside the horizon, where the maximum-volume hypersurface will saturate at late times. The new position $r_*$ of the maximum is given as in Eq. \ref{eq:rstar}, with $r_1$ given by \eq{r_1 =\frac{\sqrt{\frac{\rads \left(\sqrt{32 \mu +9 \rads^2}-3 \rads\right)}{64 \mu +18 \rads^2}} \left(3 \rads^3+\rads^2 \sqrt{32 \mu +9 \rads^2}+\mu  \sqrt{32 \mu +9 \rads^2}+9 \mu  \rads\right)}{4 \mu } .\label{eq:r1gb}} This is the local maximum of the volume functional promised in \cite{Stanford:2014jda}. By inspection, it is clear that $r_1>0$, so the $\a'$ correction moves the maximum closer to the horizon. Of course, this analysis is only valid for small $\a$; for large $\a$, the physics may be quite different, as indicated in \cite{Stanford:2014jda}.

From Figure \ref{fig:cvGBGR}, we can immediately see some important physics: as predicted on physical grounds in Section 1, we can see that the $\a'$ correction reduces the complexification rate! More quantitatively, we have that the $\calO(\a)$ portion of $\calV(r_*)$ is given by \eq{\calV(r_*) = -\frac{\a}{64}\sqrt{\sqrt{9 \rads^4+32 \mu  \rads^2}-3 \rads^2} \left(17 \rads^2+\sqrt{9 \rads^4+32 \mu  \rads^2}\right) \sqrt{\frac{\sqrt{9 \rads^4+32 \mu  \rads^2}}{\rads^2}+1},} which implies that the correction to the complexification rate is \eq{\cdot_V =  -\frac{\a}{64G\rads}\sqrt{\sqrt{9 \rads^4+32 \mu  \rads^2}-3 \rads^2} \left(17 \rads^2+\sqrt{9 \rads^4+32 \mu  \rads^2}\right) \sqrt{\frac{\sqrt{9 \rads^4+32 \mu  \rads^2}}{\rads^2}+1}.\label{eq:cvGB}}

\begin{figure}
\begin{center}
\begin{subfigure}[b]{.45\textwidth}
\begin{center}
\includegraphics[scale=.6]{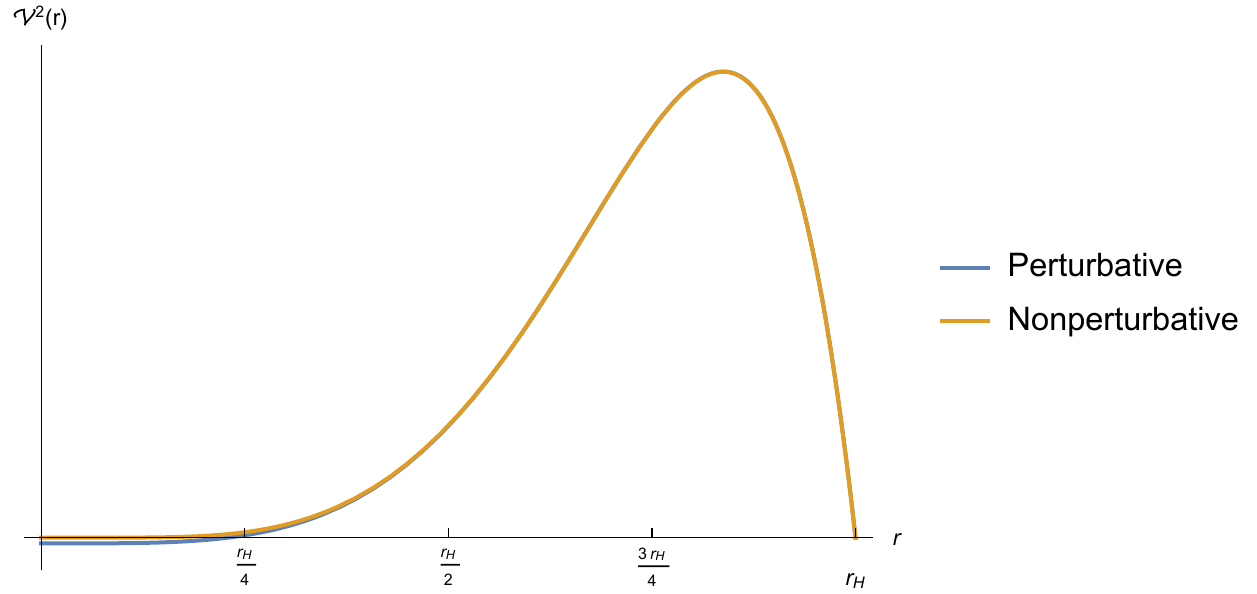}
\end{center}
\end{subfigure}
~ ~ ~ ~
\begin{subfigure}[b]{.45\textwidth}
\begin{center}
\includegraphics[scale=.6]{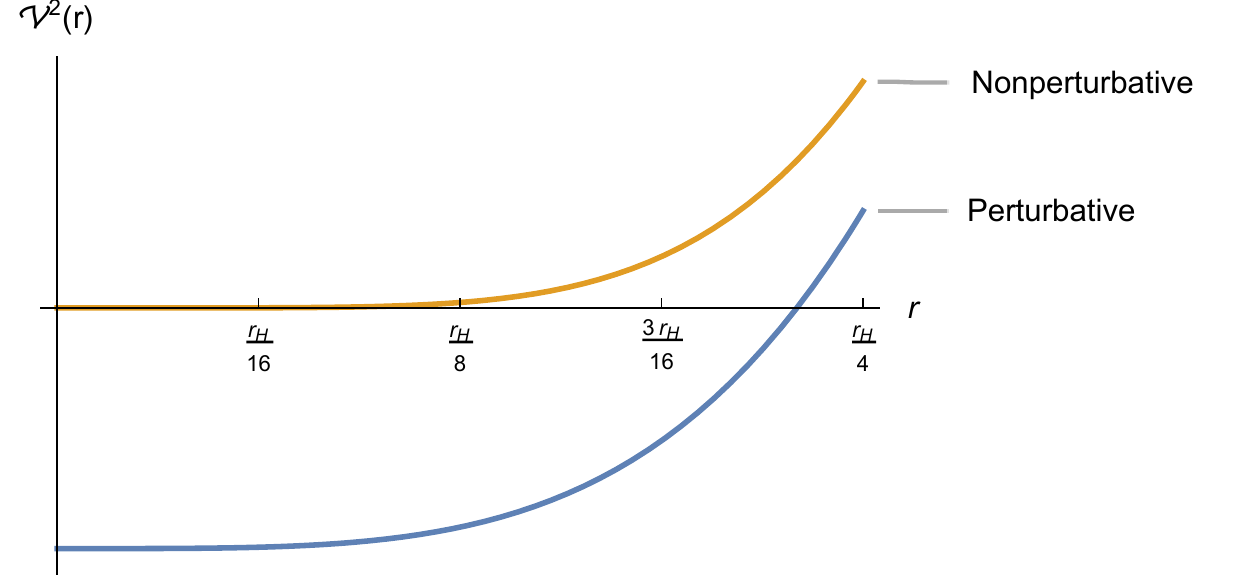}
\end{center}
\end{subfigure}
\caption{Simultaneous plots of $\calV^2$ for the GB metrics in Eq. \ref{eq:GBf1f2} and \ref{eq:gbMetricNonlinear}. We can see that the apparent negativity of $\calV^2$ in the linearized metric goes away upon inclusion of nonlinear effects.}
\label{fig:cvGBnon}
\end{center}
\end{figure}

It is worth remarking on one quirk of Figures \ref{fig:cvGB} and \ref{fig:cvGBGR}. Taken literally, it appears from these plots that $\calV^2$ is negative for very small $r$, and therefore that $\calV$ is imaginary. This is unphysical, and is merely an artifact of perturbation theory. Indeed, one can solve for the location of this zero exactly to find \eq{r = \frac{1}{64} \sqrt{\frac{\sqrt{\alpha  \rads^2 \left(4 \alpha  \mu +2 \mu +\alpha  \rads^2\right)}}{2 \alpha +1}+\frac{\alpha  \rads^2}{2 \alpha +1}} \sim 2^{1/4} \a^{1/4} \m^{1/4} \sqrt{\rads} + \calO\left(\a^{3/4}\right).} This is nonanalytic in $\a$, and should therefore not be taken seriously in a strictly perturbative framework. That higher powers in  $\a$ will erase this seeming second horizon can be seen directly from Figure \ref{fig:cvGBnon}, where we have plotted the volume functional of the nonlinear metric in Eq. \ref{eq:gbMetricNonlinear}. 

\subsection{CV for the Weyl$^4$ Action}
We will now repeat the above analysis for the Weyl$^4$ theory. Here the bulk action is\cite{Banks:1998nr,Gubser:1998nz,Pawelczyk:1998pb,Galante:2013wta,Buchel:2004di,Buchel:2008ae,Buchel:2008vz,Buchel:2006dg,deHaro:2003zd}\footnote{One might wonder whether we must worry about the effect of a spatially varying dilaton \cite{MYERS1987701}. However, it was argued in \cite{Myers:2008yi,Buchel:2008ae} that the dilaton only contributes to order $\g^2$, and thus that we can neglect it in this perturbative context. We thank R. Myers for discussion of this point.} \eq{S_{\text{Weyl, bulk}} = \frac{1}{16\piG}\int d^5x\sqrt{-g}\left[R-2\L + \g\rads^6\left(C^{\eta\mn\k}C_{\pi\mn\t}{C_\eta}^{\r\s\pi}{C^\t}_{\r\s\k} + \frac{1}{2}C^{\eta\k\mn}C_{\pi\t\mn}{C_\eta}^{\r\s\pi}{C^\t}_{\r\s\k}\right)\right],\label{eq:WbulkAction}} where we have defined a dimensionless expansion parameter \eq{\g \equiv \frac{1}{8}\zeta(3)\frac{\a'^3}{\rads^6}.\label{eq:gDef}} For convenience, we will occasionally refer to the quartic polynomial in Eq. \ref{eq:WbulkAction} simply as $W$. 

As before, we want to insert the metric ansatz in Eq. \ref{eq:abMetric}, with $a(r)$ and $b(r)$ as in Eq. \ref{eq:abDefEps}\footnote{Of course, here $\eps$ should be replaced with $\g$.}, into this action to obtain an effective Lagrangian for $f_1$ and $f_2$, exactly as was done for topological black holes in \cite{Galante:2013wta}. Doing so yields the E-L equations {\small \subeqs{0&=  3\frac{-f_2'r^{13} \left(r^4+r^2 \rads^2-\mu  \rads^2\right)-2 r^{14} f_2 \left(2 r^2+\rads^2\right)-{20 \mu ^3 \rads^6 \left(144 r^4+160 r^2 \rads^2-171 \mu  \rads^2\right)}}{2 \rads^2r^{13}}+O\left(\gamma\right)\\ 0&= 2\frac{f_1' r^{13}\left(r^4+r^2 \rads^2-\mu  \rads^2\right)+2 r^{14} f_2 \left(2 r^2+\rads^2\right)+{20 \mu ^3 \rads^8 \left(16 r^2-27 \mu \right)}}{2 \rads^2r^{13}}+O\left(\gamma\right),}} which can be solved to yield \subeqs{f_1(r) &= \frac{C_1}{r^4+r^2 \rads^2-\mu  \rads^2}+\frac{5 \mu ^3 \rads^6 \left(24 r^4+16 r^2 \rads^2-9 \mu  \rads^2\right)}{r^{12} \left(r^4+r^2 \rads^2-\mu  \rads^2\right)} + C_2\\ f_2(r) &= \frac{C_1}{r^4+r^2 \rads^2-\mu  \rads^2}+\frac{5 \mu ^3 \rads^6 \left(72 r^4+64 r^2 \rads^2-57 \mu  \rads^2\right)}{r^{12} \left(r^4+r^2 \rads^2-\mu  \rads^2\right)},} where $C_1$ and $C_2$ are integration constants. Following the same reasoning as before, we choose $C_2=0$ and pick $C_1$ so that $f_1$ and $f_2$ are well-behaved at the horizon. Doing so gives us \eq{C_1 = -5\m^3\rads^6\frac{24\rh^4+16\rh^2\rads^2-9\m \rads^2}{\rh^{12}}\label{eq:c1w}} so that  \begin{subequations}\label{eq:Wf1f2} \begin{align} f_1(r) &= \frac{5 \mu ^3 \rads^6 \left(24 r^4+16 r^2 \rads^2-9 \mu  \rads^2\right)}{r^{12} \left(r^4+r^2 \rads^2-\mu  \rads^2\right)} - \frac{5\m^3\rads^6\left(24\rh^4 + 16\rh^2\rads^2 - 9\m\rads^2\right)}{\rh^{12}\left(r^4+r^2 \rads^2-\mu  \rads^2\right)}\\ f_2(r) &= \frac{5 \mu ^3 \rads^6 \left(72 r^4+64 r^2 \rads^2-57 \mu  \rads^2\right)}{r^{12} \left(r^4+r^2 \rads^2-\mu  \rads^2\right)} - \frac{5\m^3\rads^6\left(24\rh^4 + 16\rh^2\rads^2 - 9\m\rads^2\right)}{\rh^{12}\left(r^4+r^2 \rads^2-\mu  \rads^2\right)} .\end{align}\end{subequations}

We therefore have that {\tiny \eq{\calV(r) = \sqrt{-f_0(r)}r^3\left\{1-5 \gamma\mu ^3 r^5\frac{   \left(-\frac{\mu }{r^2}+\frac{r^2}{\rads^2}+1\right) \left[\frac{\rads^{12} \left(-24 r^4-16 r^2 \rads^2+9 \mu  \rads^2\right)}{r^{12}}+\frac{960 \mu  \rads^2-256 \rads^4 \left(\sqrt{\frac{4 \mu }{\rads^2}+1}-1\right)}{\left(\sqrt{\frac{4 \mu }{\rads^2}+1}-1\right)^6}\right]}{2r^3\rads^4 \left(\rads^4 \left(r^4+r^2 \rads^2-\mu  \rads^2\right)^2\right)}\right\} + \calO\left(\g^2\right).\label{eq:calVw}} }

\begin{figure}
\begin{centering}
\includegraphics[scale=1.4]{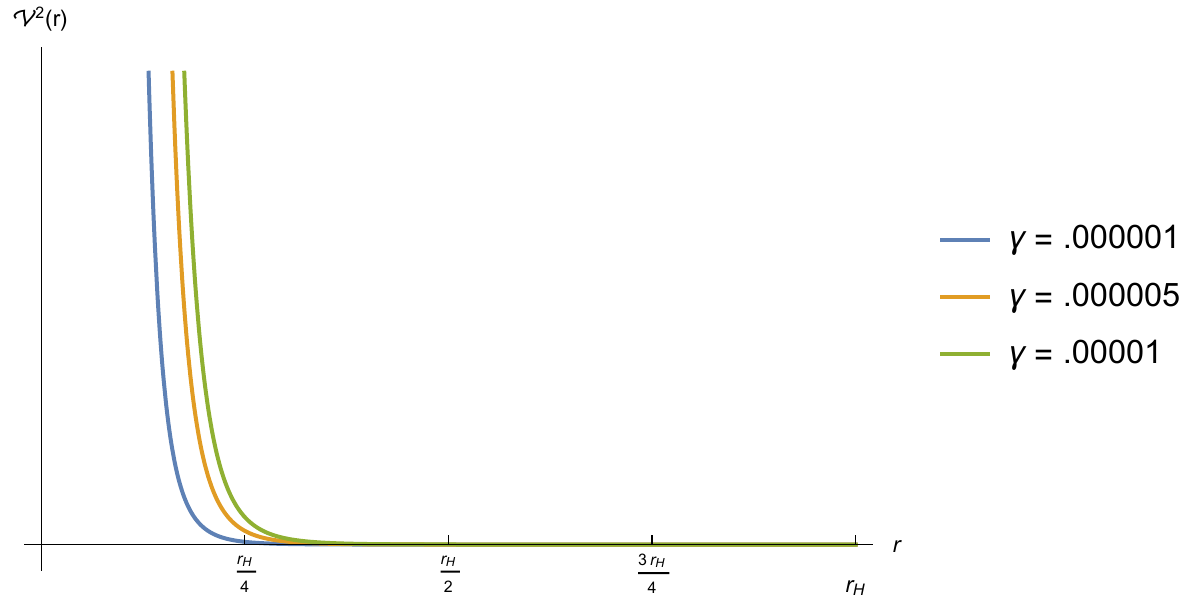}
\caption{The square $\calV^2(r)$ of the volume functional for the $\g$-corrected AdS-Schwarzschild geometry in the \wf theory, plotted for several values of $\g$ (labeled). Observe in particular that $\calV^2(r)$ diverges as $r$ approaches the singularity, in contrast to the usual limiting behavior $\calV\to0$.}
\label{fig:cvW}

\includegraphics[scale=1.4]{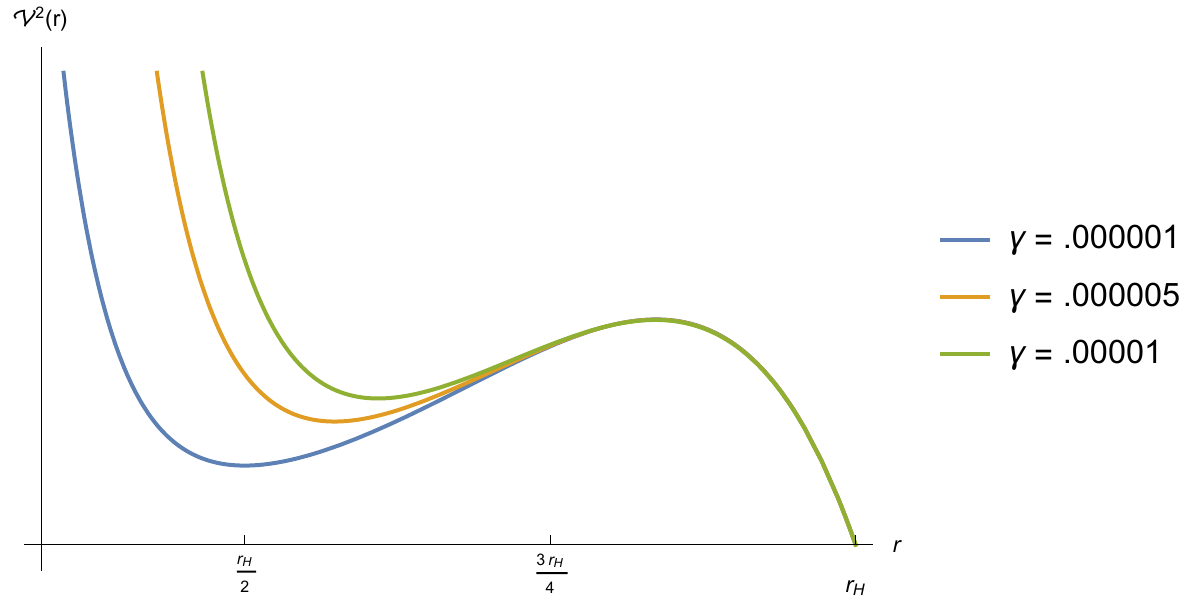}
\caption{The square $\calV^2(r)$ of the volume functional for the $\g$-corrected AdS-Schwarzschild geometry in the \wf theory, plotted for several values of $\g$ (labeled). We have adjusted the range of the $r$-axis so that the extrema are more clearly visible.}
\label{fig:cvWzoom}
\end{centering}
\end{figure}

\begin{figure}[h]
\begin{centering}
\includegraphics[scale=1.4]{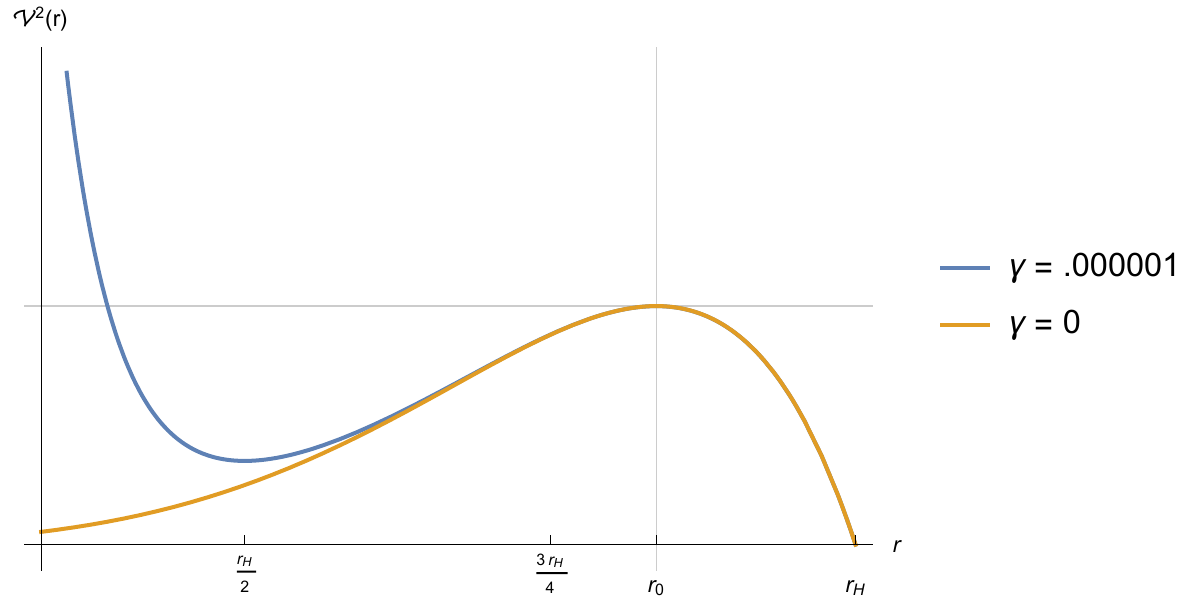}
\caption{The square $\calV^2(r)$ of the volume functional for both the Weyl$^4$-corrected AdS-Schwarzschild geometry and the uncorrected geometry. The drastic difference in behavior between the two theories is clearly visible.}
\label{fig:cvWGR}
\end{centering}
\end{figure}

We have plotted $\calV^2(r)$ for several values of $\g$ in Figures \ref{fig:cvW}. and \ref{fig:cvWzoom}. As anticipated, we see radically different physics than is traditionally observed in CV analyses; this can be seen very clearly in Figure \ref{fig:cvWGR}, where we have graphed $\calV^2(r)$ for the \wf theory alongside the pure GR curve. Instead of a single maximum inside the horizon, past which $\calV(r)$ vanishes as we approach the singularity, we now have two extrema. As we progress inwards past the usual maximum, we now hit a new minimum, beyond which  $\calV(r)$ diverges as we approach the singularity. These extrema are shown more clearly in Figure \ref{fig:cvWzoom}, where we have zoomed in on their locations. The usual caveat that our analysis is only valid for small $\g$ continues to hold here. For our usual parameter values ($\rads = 10,000$, $\m = 1,000,000,000$), numerical tests indicate that radically different behavior begins to happen around $\g\sim0.00015$; we will restrict ourselves to working at $\g$ smaller than this value. 

\begin{figure}[h]
\begin{centering}
\includegraphics[scale=1.4]{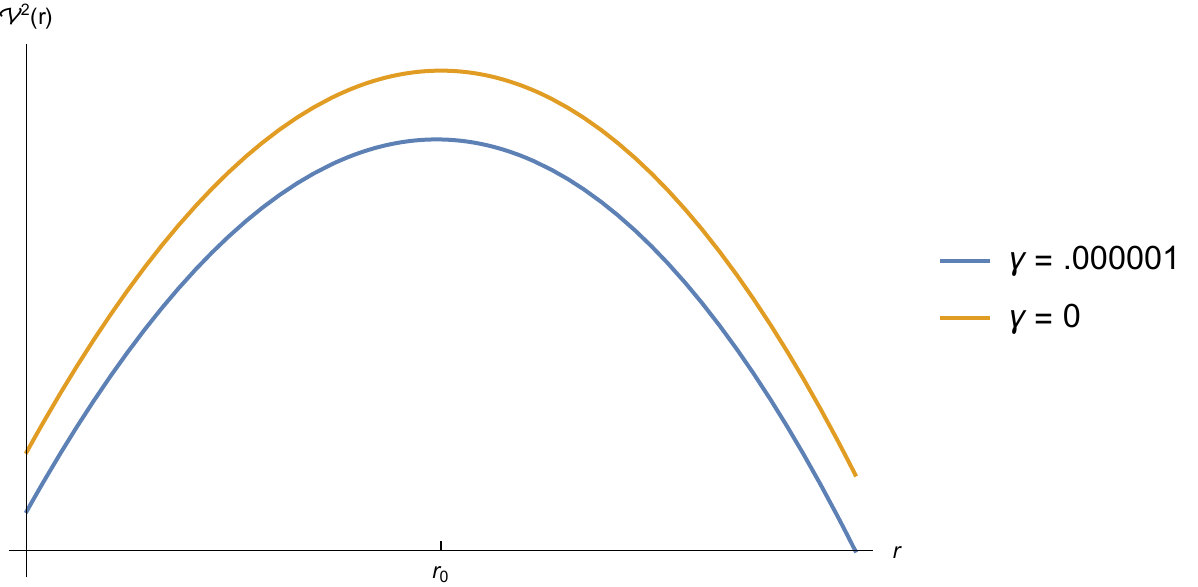}
\caption{The square $\calV^2(r)$ of the volume functional for both the Weyl$^4$-corrected AdS-Schwarzschild geometry and the uncorrected geometry. We have zoomed on on the local maximum $r_*$ to show that the correction in Eq. \ref{eq:cvW} reduces $\cdot_V$, as expected.}
\label{fig:cvWGRzoom}
\end{centering}
\end{figure}

The location of the maximum can again be found by means of Eq. \ref{eq:rstar}. Plugging the result into Eq. \ref{eq:calVw} gives that the correction is given by \seq{\calV(r_*) =\ &80\g\m^3\frac{\left(-3\rads+\sqrt{9\rads^2+32\m}\right)\sqrt{6\rads^2+32\m-2\rads\sqrt{9\rads^2+32\m}}}{\rads^7 \left[-16\m+\rads\left(-3\rads+\sqrt{9\rads^2+32\m}\right)\right]}\Bigg\{ \\ & \frac{1024\rads^8\left(3\rads^2+12\m-\rads\sqrt{9\rads^2+32\m}\right)}{\left(-3\rads+\sqrt{9\rads^2+32\m}\right)^6} + \frac{\rads^2\left[-15\m+4\rads\left(-\rads+\sqrt{\rads^2+4\m}\right)\right]}{\left(-1+\sqrt{1+\frac{4\m}{\rads^2}}\right)^6} \Bigg\}.\numberthis\label{eq:cvrsW}} It can be easily verified numerically that this correction is negative for a wide range of parameters, matching the intuition laid out in Section 1; this can also be seen in Figure \ref{fig:cvWGRzoom}, where we have zoomed in on the local maximum.  

This suggests that the correction to the complexification rate should be \seq{\cdot_V =\ &80\g\m^3\frac{\left(-3\rads+\sqrt{9\rads^2+32\m}\right)\sqrt{6\rads^2+32\m-2\rads\sqrt{9\rads^2+32\m}}}{G\rads^8 \left[-16\m+\rads\left(-3\rads+\sqrt{9\rads^2+32\m}\right)\right]}\Bigg\{ \\ & \frac{1024\rads^8\left(3\rads^2+12\m-\rads\sqrt{9\rads^2+32\m}\right)}{\left(-3\rads+\sqrt{9\rads^2+32\m}\right)^6} + \frac{\rads^2\left[-15\m+4\rads\left(-\rads+\sqrt{\rads^2+4\m}\right)\right]}{\left(-1+\sqrt{1+\frac{4\m}{\rads^2}}\right)^6} \Bigg\}.\numberthis\label{eq:cvW}} However, the interpretation of these results is somewhat subtle. Because $\calV$ diverges as we approach the singularity, the volume in Eq. \ref{eq:cvrsW} does not maximize $\calV(r)$ behind the horizion. In particular, a UV cutoff at $r=\eps$ can easily be made to satisfy \eq{\calV(\eps) > \calV(r_*)} by simply picking $\eps$ small enough. Thus, as $\eps$ approaches the singularity, the maximum of $\calV(r)$ on the interval $r\in[\eps,r_H]$ will in general be $\calV(\eps)$, which would suggest that instead of Eq. \ref{eq:cvW} we should have instead \eq{\cdot_V = \frac{\Om_3}{G\rads}\calV(\eps),\label{eq:cvWeps}} where $\calV(r)$ is given for general $r$ in Eq. \ref{eq:calVw}.

 On the other hand, it is hard to understand how we might claim to have good perturbative control over such a UV cutoff. It is implicit in Eq. \ref{eq:rstar} that $r_*$ is the only extremum of $\calV(r)$ that can be reached perturbatively. Put another way, there is no obvious way to impose a UV cutoff at $r=\eps$ in such a way that we can smoothly obtain the pure GR result in Eqs. \ref{eq:r0} and \ref{eq:cvGR} by taking a $\g\to0$ limit. Indeed, in a perturbative framework, it does not appear to be possible to even find a closed-form expression for the position of the minimum of $\calV(r)$. One could imagine solving the condition $\calV'(r)=0$ for a general $r$, instead of for $r_*$, but this amounts to solving a sixteenth-order polynomial for $r$, which of course can't be done in closed form\footnote{Numerical tests indicate that, of the 16 roots, only four of them are real: the maximum and minimum seen in Figure \ref{fig:cvWzoom} and their unphysical counterparts at negative $r$. The other twelve roots are complex, and therefore also unphysical.}. The upshot is that, working strictly in perturbation theory, it is difficult to claim to have analytical control of what happens at $r<r_*$. This is intimately related to the breakdown of gravitational EFT discussed in Section 1.

In light of the above discussion, it seems clear to use that the correct for the CV complexification rate is the finite result given in Eq. \ref{eq:cvW} rather than the divergent result in Eq. \ref{eq:cvWeps}. We will, however, return to this discussion in Section 5, after encountering similar divergences in the CA complexification rate.

\section{CA for Gauss-Bonnet Gravity}
\label{sec:GB}
Having obtained above the corrected metrics for both of our perturbed theories, we can now perform the CA analysis. We will begin with the comparatively simple case of GB gravity, for which the action is \eq{S_{\text{GB}} = S_{\text{GB, bulk}} + S_{\text{GB, bdy}},\label{eq:sGB}} where the bulk action is given in Eq. \ref{eq:5dGBbulkAction} and the complete boundary action for a constant-$r$ hypersurface is \cite{PhysRevD.36.392}  \eq{S_{\text{GB, bdy}}[r] = \frac{1}{8\piG}\int d^4x\sqrt{-h}\left\{K - 2\a\rads^2\left[ 2\calG_{ab}K^{ab} + \frac{1}{3}\left(K^3 - 3 KK_{ab}K^{ab} + 2K_a^bK_b^cK_c^a\right)\right]\right\}. \label{eq:5dGBbdyAction}} In this expression, $h_{ab}$ is the induced metric on the boundary, $\calG_{ab}$ is its Einstein tensor, and $K_{ab}$ is the extrinsic curvature tensor.

In principle, the path forward is quite simple. We simply plug the metric in Eq. \ref{eq:abMetric}, with the form of $f_1$ and $f_2$ given in Eq. \ref{eq:GBf1f2}, into Eq. \ref{eq:sGB}, expand all of the terms to linear order in $\a$, and evaluate. There are two types of contributions: one type originates from evaluating the ordinary Einstein-Hilbert action, plus the GHY term, on the perturbed metric, and the other comes from plugging the unperturbed metric itself into the GB action (and its boundary term). These two contributions are separately divergent, and the divergences do not in general cancel to give a finite answer.  

Because of the detailed nature of the calculations, we will not be especially explicit, especially with the boundary terms. To provide a more explicit example, we will first work out in Section \ref{sec:gb4} the very simple case of four-dimensional GB gravity. In four dimensions the GB term is topological, so neither the equations of motion nor their solution are corrected. Thus, of the two classes of contributions to the CA rate we discussed above, we will only need to consider the second type in this example. In Section \ref{sec:gb5} we will return to the five dimensional example. 

Before we proceed, we must mention that a similar analysis of the CA conjecture for GB gravity was performed in \cite{Cai:2016xho}, and in \cite{Cano:2018aqi} CA was studied for a generalization of GB gravity known as Lovelock gravity. Our analysis differs from those in two crucial aspects. First, these works used the nonlinear form of the metric given in Eq. \ref{eq:gbMetricNonlinear}. As mentioned above, this form of the metric is unsuited for use in perturbation theory, as it corresponds to an infinite tower of $\a'$-suppressed terms. Thus, the metric used here is more appropriate for use in a stringy context. More importantly, in both works, but especially in \cite{Cano:2018aqi}, the focus was on charged black holes. For a charged black hole geometry, the WDW patch does not extend to the singularity. Because the singularity is the origin of the divergences we encountered above, and will encounter again below, for a charged black hole these divergences cannot appear in the first place. However, given the privileged role of the TFD geometry, which corresponds to a neutral black hole, the effect of the singularity is crucial for the holographic interpretation of complexity.

\subsection{A Topological Example: GB Gravity in Four Dimensions}
\label{sec:gb4}
We begin by considering the GB action in four dimensions, for which the bulk and boundary actions are \subeqs{S_{\text{GB, bulk}} &= \frac{1}{16\piG}\int d^4x\sqrt{-g}\left[R-2\L + \a'\left(R^\mnrs R_\mnrs - 4R_\mn R^\mn + R^2\right)\right] \label{eq:GB4actionBulk} \\ S_{\text{GB, bdy}} &= \frac{1}{8\piG}\int d^3x\sqrt{-h}\left\{K - 2\a'\left[2\calG_{ab}K^{ab} + \frac{1}{3}\left(K^3 - 3 KK_{ab}K^{ab} + 2K_a^bK_b^cK_c^a\right)\right]\right\}.\label{eq:GB4actionBdy}} The four-dimensional Gauss-Bonnet(-Chern) theorem says that the Euler character $\chi_4$ of a four-manifold $M$ is given by \eq{\chi_4 = \frac{1}{32\pi^2}\int_M d^4x\sqrt{-g}\left[R_\mnrs R^\mnrs -4 R_\mn R^\mn + R^2\right],} so in four dimensions GB gravity is topological, and this theory is essentially just general relativity. In particular, the Einstein field equations do not change, so the metric does not pick up a correction, and the four-dimensional AdS-Schwarzschild solution is \eq{ds^2 = -f(r)dt^2 + \frac{dr^2}{f(r)} + r^2d\Om_2^2,\label{eq:4dds2}} where the emblackening factor $f$ is simply \eq{f(r) = 1 - \frac{2GM}{r} + \frac{r^2}{\rads^2},\label{eq:4df}} with no order $\a'$ correction.

The thermodynamics of Schwarzschild-type black holes is well understood in this theory. The temperature of the black hole horizon is proportional to $f'$ evaluated at the horizon, so the temperature therefore receives no $\calO(\a')$ contribution. Conversely, the Wald entropy does get corrected; the corrected entropy $S_{\a'}$ is given by (see e.g. \cite{Jacobson:1993xs,Chatterjee:2013daa,Charles:2016wjs,Liko:2007vi,Sarkar:2010xp}) \eq{S_{\a'} = S_0\left(1+ \a'\frac{4\pi\chi}{A}\right),\label{eq:gbEnt}} where \eq{S_0 = \frac{A}{4G}} is the Bekenstein-Hawking entropy and $\chi$ is the two-dimensional Euler character of the horizon. We are interested in spherical horizons, so we have $\chi=2$.

We will use this formula to repackage the CA complexification rate for this theory in terms of the corrected entropy, but first we must compute the rate. 
This analysis is much simpler than the CA calculations that will follow, so we will be extremely explicit. There is no metric correction, so we simply need to evaluate the order-$\a'$ action in Eqs. \ref{eq:GB4actionBulk} and \ref{eq:GB4actionBdy} on the metric in Eq. \ref{eq:4dds2}. It is straightforward to compute the curvature invariants, which are given by \subeqs{R_{\mnrs}R^{\mnrs} &= f''^2+\frac{4 f'^2}{r^2}+\frac{4 \left[f-1\right]^2}{r^4} \\ R_\mn R^\mn &= \frac{r^4 f''^2+8 r^2 f'^2+8 f \left[r f'-1\right]+4 r f' \left[r^2 f''-2\right]+4 f^2+4}{2 r^4} \\ R^2 &= \frac{\left[r^2 f''+4 r f'+2 f-2\right]^2}{r^4},} where we have dropped the explicit $r$ dependence of $f$. We will follow this convention for the remainder of the paper. We therefore have that \eq{R_{\mnrs}R^{\mnrs} - 4R_\mn R^\mn + R^2 =  \frac{4 \left\{\left[f-1\right] f''+f'^2\right\}}{r^2} = \frac{1}{r^2}\frac{d}{dr}\left\{4f'\left[f-1\right]\right\}.} Thus the $\calO(\a')$ bulk action is \seq{S_{{\text{GB, bulk}}} &= \frac{\a'}{16\piG}\int d^4x\sqrt{-g} \left[R_{\mnrs}R^{\mnrs} - 4R_\mn R^\mn + R^2\right] \\ &= \frac{\a'}{16\piG}\int dt\ dr\ d\Om_2\ r^2 \frac{1}{r^2}\frac{d}{dr}\left\{4f'\left[f-1\right]\right\} \\ &= \frac{\a'\de{t}\Om_2}{16\piG}\int_\eps^{r_H} \frac{d}{dr}\left\{4f'\left[f-1\right]\right\} \numberthis \\ &= \frac{\a'\de{t}\Om_2}{4\piG}\left.\Big\{f'\left[f-1\right]\Big\}\right|_\eps^{r_H}. \numberthis } Near the singularity, the emblackening factor scales as $f\sim 1/r$, so we have a cubic UV divergence from the singularity. 

We will now move on to the boundary term, which will be seen to cancel this divergence. For a constant-$r$ hypersurface, we choose a unit normal \eq{n_\m = \left(0,f^{-1/2},0,0\right).\label{eq:4dn}} On such a surface, we have an induced metric $h_{ab}$ given by \eq{h_{ab} = \diagthree{-f}{r^2}{r^2\sin\th^2}\label{eq:4dh}.} The Einstein tensor for this metric is \eq{\calG_{ab} = \diagthree{\frac{f}{r^2}}{0}{0}.} Similarly, from the form of $n_\m$, we can straightforwardly compute that the extrinsic curvature tensor $K_{ab}$ is given by \eq{K_{ab} \equiv \frac{1}{2}\left(\nabla_an_b + \nabla_bn_a\right) = \frac{\sqrt{f}}{2}\diagthree{-f'}{r}{r\sin^2\th}.\label{eq:4dk}} 

We can therefore work out that the curvature scalars in Eq. \ref{eq:GB4actionBdy} are \subeqs{2\mathcal{G}_{ab}K^{ab} &= -\frac{f'}{r^2\sqrt{f}} \\ K^3 &= \frac{(f')^3}{8f^{3/2}} + \frac{3(f')^2}{2r\sqrt{f}} + \frac{6f'\sqrt{f}}{r^2} + \frac{8f^{3/2}}{r^3} \\ -3KK_{ab}K^{ab} &= -3\left[\frac{(f')^3}{8f^{3/2}} + \frac{f'\sqrt{f}}{r^2} + \frac{(f')^2}{2r\sqrt{f}} + \frac{4f^{3/2}}{r^3}\right]\\ 2K_a^bK_b^cK_c^a &= \frac{(f')^3}{4f^{3/2}} + \frac{4f^{3/2}}{r^3}} so that \eq{2K_{ab}\calG^{ab} + \frac{1}{3}\left(K^3 - 3KK_{ab}K^{ab} + 2K_a^bK_b^cK_c^a\right) = \frac{f'\left(f-1\right)}{r^2\sqrt{f}}.} Thus \eq{S_{\text{GB, bdy}}[r] = -\frac{\a'}{4\pi G} \int d^3x\sqrt{-h}\left[\frac{f'(f-1)}{r^2\sqrt{f}}\right] = -\frac{\a'\Om_2\de{t}}{4\pi G} r^2\sqrt{f} \frac{f'\left(f-1\right)}{r^2\sqrt{f}} = -\frac{\a'\de{t}\Om_2}{4\piG} f'(f-1).} This is exactly the negative of the bulk action, so we have a matching cubic divergence from the $r=\eps$ boundary term that will render the total action finite. 

Putting it all together, we have that \seq{S_{\de\text{WdW}}\Big|_{\calO(\a')} &= S_{\text{GB, bulk}} + S_{\text{GB, bdy}}[r_H] - S_{\text{GB, bdy}}[\eps]\\ &= \frac{\a'\de{t}\Om_2}{4\piG}\left.\Big\{f'\left[f-1\right]\Big\}\right|_\eps^{r_H} - \frac{\a'\de{t}\Om_2}{4\piG}\left.\Big\{f'\left[f-1\right]\Big\}\right|_\eps^{r_H} = 0. \numberthis} Thus we see that, for the topological Gauss-Bonnet theory, the CA complexification rate picks up no $\calO(\a')$ correction. We therefore find that the complexification rate is exactly what we found in Appendix A, namely \eq{\cdot_A = \frac{1}{\pi}S_0T = \frac{2M}{\pi}.} However, in the corrected theory, it is more natural to write this expression in terms of the corrected quantities, so we invert Eq. \ref{eq:gbEnt} to find that \eq{\cdot_A = \frac{1}{\pi}S_{\a'}T\left(1-\a'\frac{4\pi\chi}{A}\right).}

The point of this exercise is twofold. First, we have shown an explicit example of a CA analysis in higher curvature gravity. This will enable us to be rather less explicit in the following sections, where the calculations are somewhat more involved. Much more importantly, we have seen that divergences are bound to appear in the action when higher curvature terms are allowed in the gravitational action. These divergences originate at the singularity, as did the divergence seen in the CV analysis of the \wf theory, and appear to be an essential feature of the WdW patch of a neutral black hole. In this example, all the divergences cancelled. However, this is a somewhat fine-tuned condition, and we will see below that such cancellations are not generic.

\subsection{A Dynamical Example: GB Gravity in Five Dimensions}
\label{sec:gb5}
We now turn to the five dimensional GB theory. As mentioned above, we will be rather less explicit here than we were in the previous section. To recap briefly, the action is given in Eqs. \ref{eq:5dGBbulkAction} and \ref{eq:5dGBbdyAction}, and the metric is of the form \eq{ds^2 = -a(r)dt^2 + \frac{dr^2}{a(r)} + r^2d\Om_3^2,} where $a(r)$ is defined in terms of $f_0(r)$ and $f_1(r)$ as in Eq. \ref{eq:abDefEps} and $f_1(r)=f_2(r)$ is given in Eq. \ref{eq:GBf1f2}. These are all of the ingredients for this calculation, and now all that remains is to assemble them. 

We will begin with the bulk terms. For a metric of the form given in Eq. \ref{eq:abMetric}, it is straightforward to verify that \begin{subequations}\label{eq:gbBulkInvts} \begin{align} 
R &= -\frac{a''r^2 + 6 \left(r a'+a-1\right)}{r^2} \\
R_\mnrs R^\mnrs &= \frac{a''^2r^4+6 r^2 a'^2+12 (a-1)^2}{r^4}\\
R_{\mn}R^\mn &=\frac{a''^2r^4+3 \left(5 r^2 a'^2+8 r (a-1) a'+8 (a-1)^2\right)+6a'a''r^3}{2r^4}\\ 
R^2 &= \frac{\left[a''r^2 + 6 \left(r a'+a-1\right)\right]^2}{r^4}.\end{align}\end{subequations} These curvature invariants depend only on $r$, so the time and angular integrals in Eq. \ref{eq:5dGBbulkAction} are trivial. We can therefore evaluate them straightforwardly to find {\small \eq{S_{\text{GB, bulk}} &= \frac{\Om_3\de{t}}{16\piG}\int_\eps^\rh dr\ r^3\left[R - 2\L + \a\rads^2\left(R_\mnrs R^\mnrs - 4R_\mn R^\mn + R^2\right)\right]\\ &= \frac{\Om_3\de{t}}{16\piG}\int_\eps^\rh dr\ \left\{ \left[-a''r^3 -6 r^2 a'-6 r a+6 r\right] +\alpha\rads^2  \left[-12 r a''+12 r a a''+12 r a'^2+24 a a'-24 a'\right]\right\}.}} Although only the second term has an explicit $\a$ factor, both terms have a $\calO(\a)$ contribution, as can readily be seen from Eq. \ref{eq:abDefEps}. Inserting the coordinate form of $a(r)$ in Eq. \ref{eq:GBf1f2} and discarding the $\calO(\a^0)$ term from the unperturbed Ricci scalar, we have \eq{S_{\text{GB, bulk}} &= \frac{\Om_3\de{t}\a}{16\piG}\int_\eps^\rh dr\ \left[\frac{80r^3}{\rads^2} + \frac{48\rads^2\m^2}{r^5}\right] \\ &= \frac{\Om_3\de{t}\a}{16\piG} \left.\left[\frac{20r^4}{\rads^2}-\frac{12\rads^2\m^2}{r^4}\right]\right|_\eps^\rh.\label{eq:GBbulkFinal}}

This is the complete bulk contribution to the CA complexification rate for the 5d GB theory. We will now proceed to the boundary terms. The only boundaries we are interested in are constant-$r$ hypersurfaces. For such boundary surfaces, we can define a purely radial normal vector and from there compute the extrinsic curvature matrix exactly as was done before. This gives us \begin{subequations}\label{eq:gbBdyInvts} \begin{align} 
K &= \frac{\sqrt{a}}{2}\left(\frac{a'}{a}+\frac{6}{r}\right)\\ 
K_{ab}\calG^{ab} &= -\frac{3 \left(r a'+2 a\right)}{2 r^3 \sqrt{a}}\\ 
K^3 &= \frac{a^{3/2}}{8}\left(\frac{a'}{a}+\frac{6}{r}\right)^3\\ 
KK_{ab}K^{ab} &= \frac{\left(r a'+6 a\right) \left(r^2 a'^2+12 a^2\right)}{8 r^3 a^{3/2}}\\
K_a^bK_b^cK_c^a &= \frac{a'^3r^3+24a^3}{8 a^{3/2}r^3}.\end{align}\end{subequations}

As before, these invariants depend only on $r$, so the time and angular integrals are trivial, leaving us with \eq{S_{\text{GB, bdy}}[r] &= \frac{\Om_3\de{t}}{8\piG}r^3\sqrt{a}\left\{K - 2\a\rads^2\left[ 2\calG_{ab}K^{ab} + \frac{1}{3}\left(K^3 - 3 KK_{ab}K^{ab} + 2K_a^bK_b^cK_c^a\right)\right]\right\}\\ &= \frac{\Om_3\de{t}}{8\piG}\left[\left(\frac{1}{2} r^3 a'+3 r^2 a\right)+\alpha\rads^2  \left(6 r  a'-6 r  a a'-4 a^2+12  a\right)\right].} As before, we now plug in the coordinate expression for $a$ and drop the $\calO(\a^0)$ term to find that \eq{S_{\text{GB, bdy}}[r] &= \frac{\Om_3\de{t}\a}{8\piG}\left[8 \mu +\frac{8 \mu ^2 \rads^2}{r^4}-\frac{8 r^4}{\rads^2}-\frac{4 \mu  \rads^2}{r^2}+4 r^2-\frac{4 \mu ^2 \rads^2}{\rh^4}-\frac{4 \rh^4}{\rads^2}+8 \rads^2\right].\label{eq:GBbdyFinal}}

We now have all of the components we need. Plugging Eqs. \ref{eq:GBbulkFinal} and \ref{eq:GBbdyFinal} into Eq. \ref{eq:SWdW}, we have \eq{S_{\de\text{WdW}} &= \left.\frac{\Om_3\de{t}\a}{\piG}\left[\frac{\mu ^2 \rads^2}{4 r^4}+\frac{r^4}{4 \rads^2}-\frac{\mu  \rads^2}{2 r^2}+\frac{r^2}{2}-\frac{\rads^4 \left(\mu ^2-2 \rh^4\right)-2 \mu  \rads^2 \rh^4+\rh^8}{2 \left(\rads^2 \rh^4\right)}\right]\right|_\eps^\rh.} We can then insert this into Eq. \ref{eq:CA} to find the correction to the CA complexification rate for 5d Gauss-Bonnet gravity: \eq{\cdot_A = \left.\frac{\Om_3\a}{\pi^2G}\left[\frac{\mu ^2 \rads^2}{4 r^4}+\frac{r^4}{4 \rads^2}-\frac{\mu  \rads^2}{2 r^2}+\frac{r^2}{2}\right]\right|_\eps^\rh.\label{eq:CVGB}} 

Eq. \ref{eq:CVGB} is the main result of this section. In particular, it contains a divergent part: \eq{\cdot_A^{\text{divergent}} = -\frac{\Om_3\a}{\pi^2G}\left[\frac{\m^2\rads^2}{4\eps^4} - \frac{\m\rads^2}{2\eps^2}\right].} This divergence is troubling, and requires careful interpretation. Before we interpret this result, however, we will see in the next section that a similar result holds in the \wf theory.

\section{CA for the Weyl$^4$ Action}
\label{sec:W4}
We will now move on to the CA analysis for the Weyl$^4$ theory. In the context of the CA conjecture, this calculation gives the leading finite-$\l$ correction to the rate of complexification for $\calN=4$ SYM in the 't Hooft limit. As before, we have \eq{S_{\text{Weyl}} = S_{\text{Weyl, bulk}} + S_{\text{Weyl, bdy}}.} We will begin with the bulk part, shown in Eq. \ref{eq:WbulkAction}, and proceed exactly as we did above. Inserting Eqs. \ref{eq:abMetric} and \ref{eq:Wf1f2} into $S_{\text{Weyl, bulk}}$ gives \eq{S_{\text{Weyl, bulk}} &= \frac{\Om_3\de{t}}{16\piG} \int_\eps^\rh dr\ r^3\sqrt{ab} \left[R - 2\L + \g\rads^6W\right] \\ &= \frac{\Om_3\de{t}}{16\piG} \int_\eps^\rh dr\left[\frac{8 r^3}{\rads^2} + \g\left(\frac{360 \mu ^4 \rads^6}{r^{13}}+\frac{960 \mu ^3 \rads^4}{r^9}\right)+O\left(\gamma ^2\right)\right].} Dropping the $\calO(\g^0)$ term and integrating, we have that \eq{S_{\text{Weyl, bulk}} = \frac{\Om_3\de{t}\g}{16\piG}\left.\left[-\frac{120\rads^4\m^3}{r^8} - \frac{30\rads^6\m^4}{r^{12}}\right]\right|_\eps^\rh\label{eq:WbulkFinal}.}

We are now free to move on to the boundary term. Although a boundary action appropriate for some propagating metric modes has been introduced in the context of $\calN=4$ hydrodynamics \cite{Buchel:2004di}, to the best of our knowledge no explicit boundary term for the Weyl$^4$ theory has been given in the literature. This stands in stark contrast with the GB theory considered above, for which the appropriate boundary term is well-known \cite{PhysRevD.36.392}. The crucial difference is that, whereas Gauss-Bonnet gravity is not a higher derivative theory, the same is not true of the \wf theory. Indeed, general $f(\text{Riemann})$ theories have fourth order equations of motion, and the \wf theory is no exception. For general higher-derivative theories, the question of well-posedness is subtle (see e.g. \cite{Simon:1990ic,Dyer:2008hb}), and it is not a priori clear that there should exist a simple boundary term. 

However, a boundary term for $f(\text{Riemann})$ theories was recently derived in \cite{Deruelle:2009zk,Jiang:2018sqj}. We will briefly describe the origin of this boundary term before evaluating it. We begin with a generic $f(\text{Riemann})$ action, \eq{S = \frac{1}{16\piG}\int d^dx\sqrt{-g} f\left(R_\mnrs\right),} and introduce auxiliary fields $\phi_\mnrs$ and $\psi_\mnrs$, which we take to have the same symmetries as the Riemann tensor, so that $\psi_\mnrs = \psi_{\r\s\mn},$ etc. We repackage the action in terms of these fields as \eq{S = \frac{1}{16\piG} \int d^dx\sqrt{-g}\left[f\left(\phi_\mnrs\right) - \psi^\mnrs\left(\phi_\mnrs-R_\mnrs\right)\right].} We see immediately that $\psi_\mnrs$ is simply a Lagrange multiplier setting $\phi$ equal to the Riemann tensor on shell. This action has equations of motion  \cite{Deruelle:2009zk,Jiang:2018sqj} {\small \subeqs{8\piG T_\mn &= -{R^{(\m}}_{\a\b\g}\psi^{\n)\a\b\g} - 2\nabla_\a\nabla_\b\psi^{\a(\mn)\b}+2{\phi^{(\m}}_{\a\b\g}\frac{\d f}{\d\phi_{\n)\a\b\g}} - \frac{1}{2}g^\mn\left[\psi^{\a\b\g\de}\left(R_{\a\b\g\de}-\phi_{\a\b\g\de}\right) + f\left(\phi_{\a\b\g\de}\right)\right] \label{eq:frEOM}\\ \psi^\mnrs &= \frac{\d f}{\d \phi_\mnrs}\label{eq:psiEOM} \\ \phi^\mnrs &= R^\mnrs\label{eq:phiEOM}.}}

If we insert Eqs. \ref{eq:psiEOM} and \ref{eq:phiEOM} into Eq. \ref{eq:frEOM}, we recover the well-known equations of motion for $f(\text{Riemann})$ gravity. The crucial point, however, is that, at the level of the equations of motion, this is not a higher-derivative theory: nowhere do we find more than two derivatives acting on any single field. Thus, we expect to be able to find a boundary term. Such a term was indeed found, and for spacelike hypersurfaces takes the form \cite{Deruelle:2009zk,Jiang:2018sqj}\eq{S_{\text{bdy}} = \frac{1}{4\piG}\int d^{d-1}x\sqrt{-h} \psi^\mnrs K_{\m\r}n_\n n_\s,} where $n_\m$ is the outward pointing unit normal. Inserting Eq. \ref{eq:psiEOM}, we have \eq{S_{\text{bdy}} = \frac{1}{4\piG}\int d^{d-1}x\sqrt{-h} \frac{\d f}{\d R_\mnrs} K_{\m\r}n_\n n_\s.}

\begin{figure}[h]
\begin{centering}
\includegraphics[scale=.5]{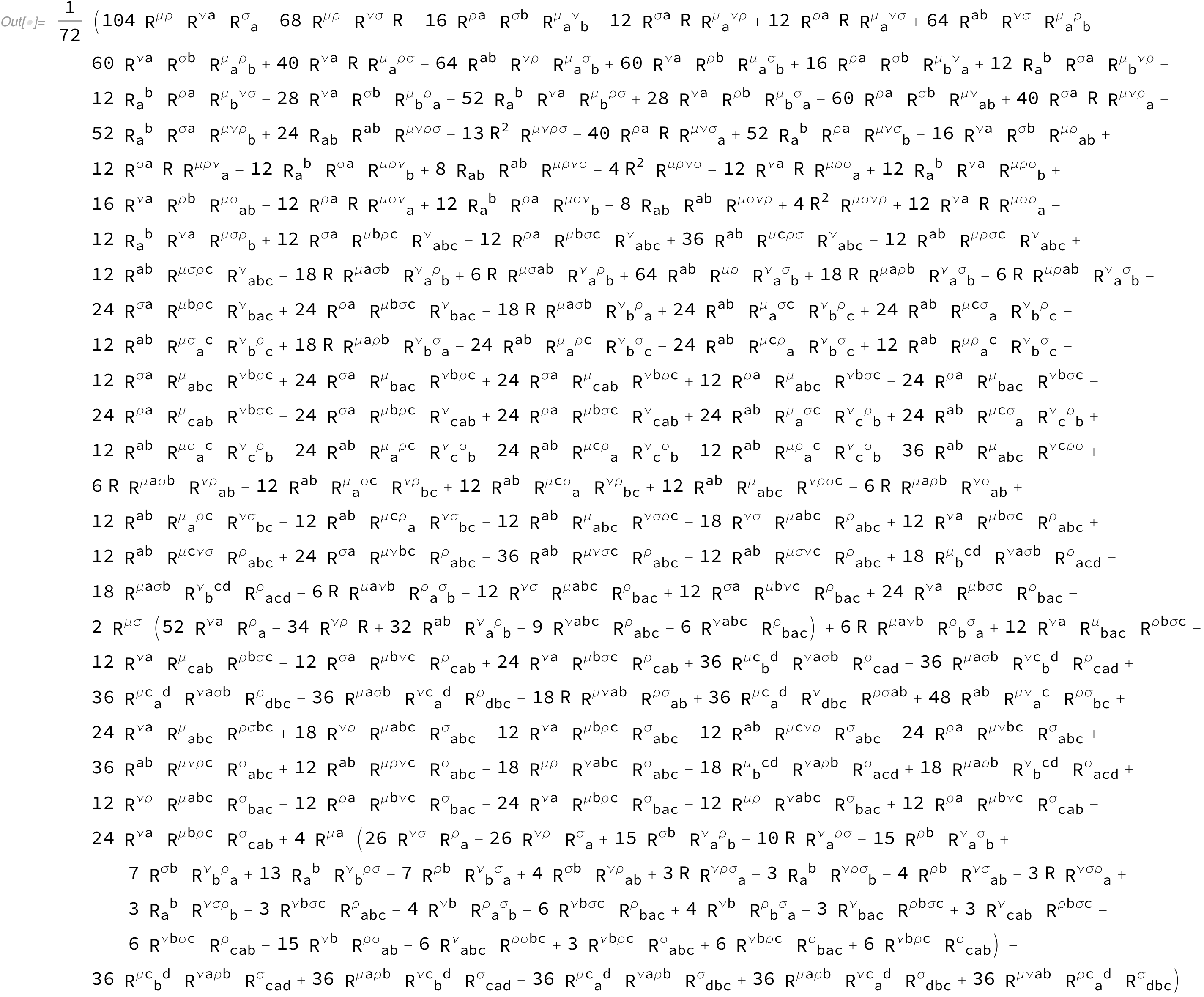}
\caption{The closed form of $dW/dR_\mnrs$. Internal indices are given Latin labels.}
\label{fig:dWdRiem}
\end{centering}
\end{figure}

In the case of interest, we have \eq{f\left(R_\mnrs\right) = W,} so we are led to consider the following boundary term: \eq{S_{\text{Weyl, bdy}} = \frac{1}{8\piG}\int d^4x\sqrt{-h}\left[K + 2\g \frac{\d W}{R_{\mnrs}} K_{\m\r}n_{\n}n_\s\right].\label{eq:WbdyAction}} In principle, all we need to do is insert the metric in Eq. \ref{eq:abMetric} into this action, but that requires us to actually evaluate the derivative in Eq. \ref{eq:WbdyAction}. The most convenient way to do so is to use the definition \eq{C_{\mu\nu\rho\sigma} = R_{\mu\nu\rho\sigma}-\frac{2}{D-2}\Big\{g_{\mu[\rho}R_{\sigma]\nu}-g_{\nu[\rho}R_{\sigma]\mu}\Big\}+\frac{2}{(D-2)(D-1)}\Bigg\{g_{\mu[\rho}g_{\sigma]\nu}R\Bigg\}} to express everything in terms of the Riemann tensor itself, and then differentiate term-by-term. The result is shown in Figure \ref{fig:dWdRiem}. 

We can now directly plug in the metric, to find the surprisingly simple result {\footnotesize \eq{S_{\text{Weyl, bdy}} &= \frac{\Om_3\de{t}\g}{8\piG}\left[\frac{2 C_1}{\rads^2}-\frac{1688 \mu }{9}+\frac{60 \mu ^4 \rads^6}{r^{12}}+\frac{60 \mu ^3 \rads^6}{r^{10}}+\frac{208 \mu ^3 \rads^4}{r^8}+\frac{64 \mu ^2 \rads^4}{r^6}+\frac{96 \mu ^2 \rads^2}{r^4}+\frac{3376 r^4}{9 \rads^2}+\frac{844 r^2}{3}\right],\label{eq:WbdyFinal}}} where as usual we have dropped the $\calO(\g^0)$ piece and $C_1$ is the integration constant in Eq. \ref{eq:c1w}; we have left it in this form because it will not effect the final answer.

We can now evaluate the complete complexification rate.  Plugging Eqs. \ref{eq:WbulkFinal} and \ref{eq:WbdyFinal} into Eq. \ref{eq:SWdW}, we have {\eq{S_{\de\text{WdW}} = \frac{\Om_3\de{t}\g}{\piG}\left.\left[\frac{45 \mu ^4 \rads^6}{8 r^{12}}+\frac{15 \mu ^3 \rads^6}{2 r^{10}}+\frac{37 \mu ^3 \rads^4}{2 r^8}+\frac{8 \mu ^2 \rads^4}{r^6}+\frac{12 \mu ^2 \rads^2}{r^4}+\frac{211 r^2}{6}+\frac{422 r^4}{9 \rads^2}\right]\right|_\eps^\rh.} } Thus we have the correction to the CA complexification rate for the \wf theory: { \eq{\cdot_A = \frac{\Om_3\g}{\pi^2G}\left.\left[\frac{45 \mu ^4 \rads^6}{8 r^{12}}+\frac{15 \mu ^3 \rads^6}{2 r^{10}}+\frac{37 \mu ^3 \rads^4}{2 r^8}+\frac{8 \mu ^2 \rads^4}{r^6}+\frac{12 \mu ^2 \rads^2}{r^4}+\frac{211 r^2}{6}+\frac{422 r^4}{9 \rads^2}\right]\right|_\eps^\rh.\label{eq:wFinal}} } Eq. \ref{eq:wFinal} is the main result of this section. As before, the complexification rate has a divergent part: \eq{\cdot_A^{\text{divergent}} = -\frac{\Om_3\g}{\pi^2G}\left[\frac{45 \mu ^4 \rads^6}{8 \eps^{12}}+\frac{15 \mu ^3 \rads^6}{2 \eps^{10}}+\frac{37 \mu ^3 \rads^4}{2 \eps^8}+\frac{8 \mu ^2 \rads^4}{\eps^6}+\frac{12 \mu ^2 \rads^2}{\eps^4}+\frac{211 \eps^2}{6}+\frac{422 \eps^4}{9 \rads^2}\right].} This divergence is manifestly negative, continuing the trend we have seen in all of our results. We will spend the remainder of the paper interpreting this divergence, as well as the other divergences that we have encountered.

\section{Conclusion}
\label{sec:conc}
We have studied holographic complexity in the Gauss-Bonnet and \wf theories and found, through a careful perturbative analysis, a series of unexpected divergences in the complexification rate, especially in the CA framework. These divergences ultimately originate from the singularity in the bulk geometry, and to the best of our knowledge have not yet been observed in the literature.

In light of the discussion in Section 1 of gravitational EFT and its breakdown at the singularity, it is perhaps somewhat unsurprising that CA calculations at higher curvature are divergent. After all, we are trusting semiclassical gravity in a regime where it shouldn't work! On the other hand, CV physics traditionally has been observed to avoid the singularity, even in the presence of higher-curvature corrections. We see here that this pattern continues, albeit with reservations. Although the volume functional $\calV(r)$ does diverge at the singularity, it is nevertheless possible to obtain a sensible perturbatively finite answer by simply neglecting the geometry near the singularity. This of course is not possible in the CA framework, where we have no choice but to go right down to the singularity. In this context, therefore, CV seems like a safer proposal, since we should always be able to avoid the high-curvature region in a way that is impossible in CA. However, even in CV there may be divergences near the singularity, and so we have learned an important lesson: whenever we are behind the horizon, as we must be to study complexity \cite{Susskind:2014moa}, we must be wary of the singularity, even when naively we might think ourselves safe.

In the CA context, our results would seem to suggest that, once stringy effects are taken into account, all neutral black holes perform infinitely fast decomplexification! This immediately seems unphysical. From the Lloyd bound \cite{2000Natur.406.1047L} and the chaos bound \cite{Maldacena:2001kr}, we expected stringy effects to lower the complexification rate relative to the case of pure general relativity. Even still, the geometries we have studied here correspond to finite temperature boundary states, and on general grounds we expect such states to complexify, rather than decomplexify. Even a zero-temperature system, such as a BPS black hole, has a vanishing complexification rate, rather than a negative rate, much less a divergent negative rate, so it is a priori unclear what this result would even mean from a strictly computational point of view. 

\begin{figure}
\begin{centering}
\includegraphics[scale=1]{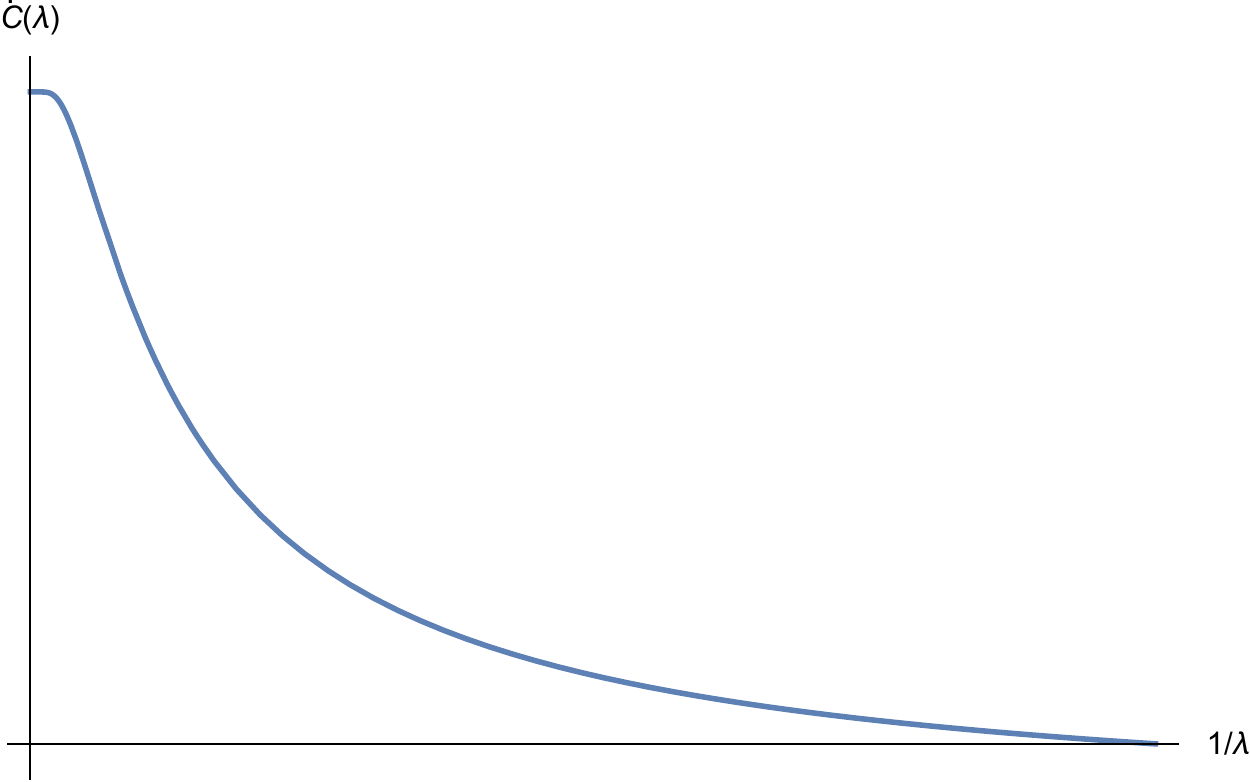}
\caption{A possible graph of $\cdot(\l)$. We have chosen this function to demonstrate visually the possible effects of $\cdot(\l)$ not being analytic at $\l=\infty$, but do not claim that it should be interpreted as a precise prediction for the form of $\cdot(\l)$.}
\label{fig:cdotl}
\end{centering}
\end{figure}

On the other hand, without a definition of the complexity of a state in the strongly coupled boundary field theory, it is hard to make this intuition precise, and a principled interpretation of these results seems beyond our reach. Nevertheless, it seems clear to us, even without a precise boundary picture, that these results should not be taken to imply that CA predicts a divergent complexification rate. Instead, we have a situation where, when viewed as a function of the 't Hooft parameter\footnote{In light of the discussion in \cite{Myers:2008yi,Buchel:2008ae}, the parameter with respect to which we have expanded is not simply $\l$ but instead a linear combination of $\l^{-3/2}$ and $\sqrt{\l}/N^2$. For brevity, we will simply refer to this as the 't Hooft coupling.}, $\cdot_A(\l=\infty)$ is well-defined, but its derivatives are infinite. This suggests to us that $\cdot_A(\l)$ is simply not analytic around $\l=\infty$\footnote{This interpretation was suggested by L. Susskind.}. In this case, the first derivative would naturally be ill-defined at $\l=\infty$, but could be well-defined elsewhere. Consider as an example the function $\sqrt{x}$. For small $x$, we could naively try expanding $\sqrt{x}$ about $x=0$ to find \eq{\sqrt{x}\sim \sqrt{0} + \frac{x}{2\sqrt{0}} \sim 0 + \infty.} Thus we see that the Taylor series doesn't converge, even though we know that $\sqrt{x}$ is well defined for all positive $x$. This seems extremely reminiscent of what we have observed here, and so we are led to conjecture that $\cdot_A(\l)$ is nonanalytic. With this in mind, we have sketched a possible graph of $\cdot_A(\l)$ in Figure \ref{fig:cdotl}. We certainly do not claim that this is an accurate depiction, but it clearly illustrates the nonanalyticity at $\l=\infty$.

This suggests a natural question: can the $\a'$ series of corrections to the complexification rate be resummed? Even if each term in a Taylor series diverges, we can occasionally extract a meaningful, finite answer from it. It is tempting to conjecture, in light of the discussion above, that this should be possible, and that by doing so we could indeed see that $\cdot_A(\l)$ is well-defined for large but finite $\l$. However, doing so would be extremely difficult: in addition to an infinite number of terms in the bulk Lagrangian, we would need to solve for an infinite number of metric corrections, as well as an infinite number of boundary terms (although in the absence of gauge field contributions the results of \cite{Jiang:2018sqj,Deruelle:2009zk} should suffice to compute many of these boundary terms). Thus, actually resumming the series seems implausible, but it might be possible to argue on general grounds that the series is resummable, and therefore that $\cdot_A(\l)$ is well-defined.

\begin{figure}
\begin{centering}
\includegraphics[scale=1.4]{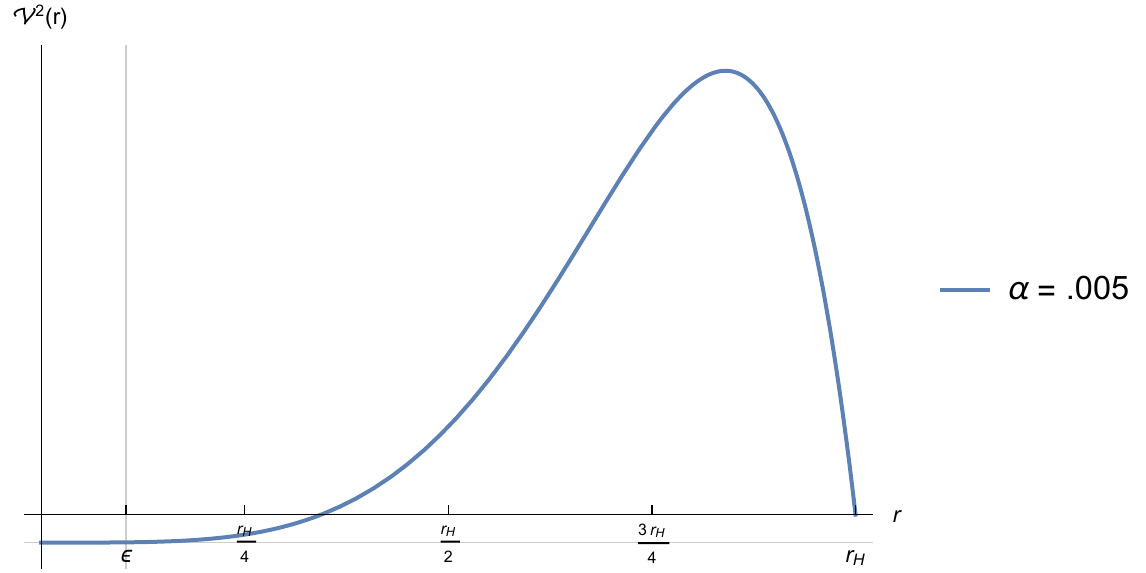}
\caption{The square $\calV^2(r)$ of the volume functional for the $\a$-corrected AdS-Schwarzschild geometry in the GB theory, with the cutoff in Eq. \ref{eq:epsDef} marked.}
\label{fig:cvGBEps}

\includegraphics[scale=1.4]{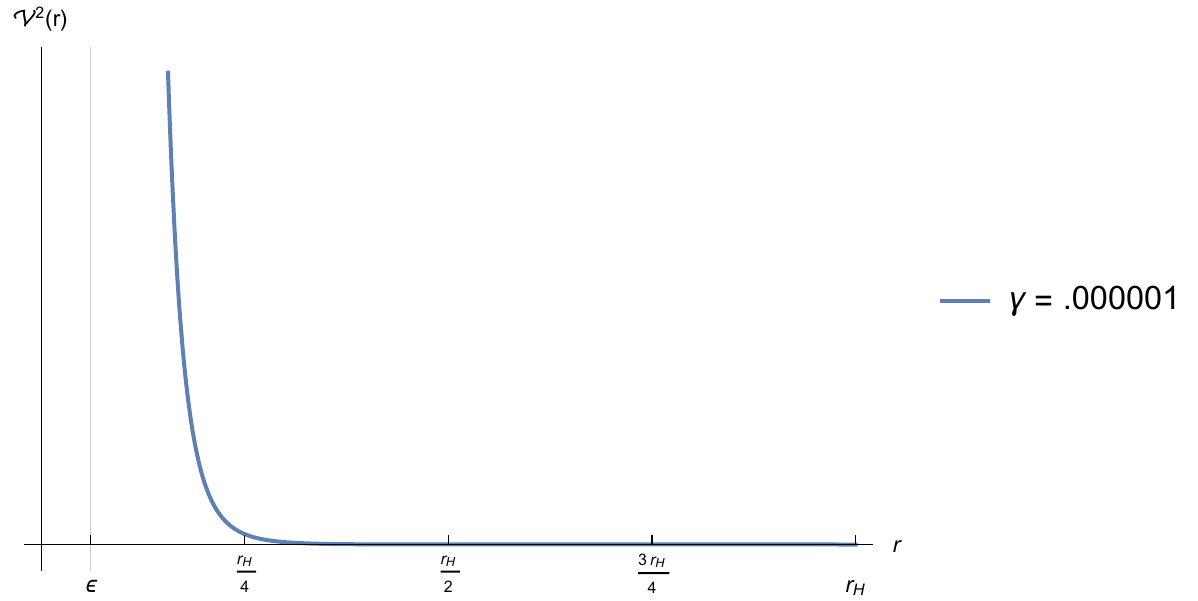}
\caption{The square $\calV^2(r)$ of the volume functional for the $\a$-corrected AdS-Schwarzschild geometry in the \wf theory, with the cutoff in Eq. \ref{eq:epsDef} marked.}
\label{fig:cvWEps}
\end{centering}
\end{figure}

Even in the absence of an explicit infinite series to be resummed, it would be interesting to obtain an estimate for the value of $\cdot_A(\l)$ at large but finite $\l$. To do so, we need a quantitative estimate of where the cutoff $\eps$ should be placed. We can choose the cutoff to be no less than a string length away from the singularity, where the curvature should be large on the string scale and our perturbation theory breaks down. Using the minimum value of the cutoff, from Eqs. \ref{eq:aDef} and \ref{eq:gDef}, we have that the cutoffs for our two theories should be located at \eq{\eps \sim \left\{ \begin{array}{cc} \a^{1/3}\rads, & \text{GB}\\ \g^{1/6}\rads, & \text{Weyl}^4 \end{array}\right..\label{eq:epsDef}} We have reproduced the plots of $\calV^2(r)$ for our two theories with $\eps$ marked in Figures \ref{fig:cvGBEps} and \ref{fig:cvWEps}. 

This simple prescription allows us to provide an estimate for the CA complexification rate for both theories. Eqs. \ref{eq:CVGB} and \ref{eq:wFinal} both contain an explicit $\eps$ dependence. We can obtain an estimate by simply plugging in the appropriate value of $\eps$ from Eq. \ref{eq:epsDef} into these expressions. For large but finite $\l$, this should give a numerically reasonable estimate for $\cdot_A(\l)$, as the portion behind the cutoff corresponds to a breakdown of perturbation theory. However, these estimates, although always finite, should not be thought of as genuine perturbative corrections. In general, they will contain fractional powers of $\l$, and therefore are not a genuine Taylor (or Laurent) series. This of course simply reflects the nonanalyticity we observed above.  

On the other hand, the interpretation of the CV results is significantly more straightforward. In Eqs. \ref{eq:cvGB} and \ref{eq:cvW}, we have finite expressions for the leading perturbative correction to the strict GR results of \cite{Susskind:2014rva,Stanford:2014jda,Susskind:2014moa}. These results suggest that the CV complexification rate $\cdot_V(\l)$ should admit a well-defined Taylor series at $\l=\infty$. This of course means that the CV framework is compatible with the complexification rate being an analytic function of the 't Hooft coupling, in stark contrast with what was seen for CA. This simpler behavior reflects that the CV physics remains in the low-curvature region, and therefore remains amenable to a perturbative treatment. From the boundary side, this lines up very nicely with the proposed boundary dual of volume in \cite{Belin:2018fxe,Belin:2018bpg}, from which we would expect that, if CV is ultimately the correct proposal, then the complexity should be an analytic function of the coupling.

In the presence of a well-defined proposal for boundary complexification rates for genuine CFTs, these results would allow us to distinguish between the CV and CA results. If hypothetically a boundary result existed, and admitted a Taylor series in the 't Hooft coupling, we could hope to discard CA in favor of CV based on these results alone. Conversely, if corrections to this boundary calculation also diverged, we could hope to attain precision matching of the coefficients and orders of the divergences in the boundary and CA calculations. Our results would therefore allow us to differentiate the CA and CV conjectures from each other in a way that is impossible in the strict GR limit\footnote{In the context of AdS3/CFT2, it is possible to distinguish between CV and CA, even at the level of pure GR, by perturbing the theory with a small, local conformal transformation \cite{Flory:2018akz,Flory:2019kah}. However, for CFTs above two dimensions, and therefore in particular for $\calN=4$ SYM, such local transformations do not exist, and thus $a\ priori$ these results to not apply.}. Although the search for a precise definition of complexity for boundary states is a hot topic for current research (see e.g. \cite{Chapman:2017rqy,Jefferson:2017sdb,Khan:2018rzm,Hackl:2018ptj,Chapman:2018hou,Guo:2018kzl,Yang:2017nfn,Yang:2018tpo,Jiang:2018nzg,Parker:2018yvk,Bhattacharyya:2018wym,Bhattacharyya:2018bbv,Ali:2018fcz,Lin:2018cbk,Belin:2018fxe,Belin:2018bpg}), it remains an open question, and it is not clear that we should expect a boundary comparison in the near future.

\addcontentsline{toc}{section}{Acknowledgements}
\section*{Acknowledgements}
R.N. is funded by the Stanford University Physics Department, a Stanford University Enhancing Diversity in Graduate Eduction (EDGE) grant, and by NSF Fellowship number DGE-1656518. He thanks A. Abdel-Aziz, R. Hennigar, H. Marrochio, P. Saad, J. Sorce, and A. Wall, and especially J. Maltz and L. Susskind, for useful discussions. He is especially grateful to A. Brown, R. Myers, and Y. Zhao for comments on a draft of this paper, and to A. Lewkowycz for suggesting the Weyl$^4$ theory.

\appendices
\section{Review of Holographic Complexity}
\label{sec:review}
In the main body of the paper, we will analyze the CA and CV conjectures in the context of higher-curvature gravity. We will in this Appendix perform the corresponding calculations in the much simpler context of Einstein gravity. This will serve to establish the notation, some of which is nonstandard, used in the rest of the paper. Throughout, we will specialize to five dimensions, which is the main case of interest in the main text. In five dimensions, the AdS-Schwarzschild solution for pure Einstein gravity is given by \eq{ds^2 = -f(r)dt^2 + \frac{dr^2}{f(r)} + r^2d\Om_3^2,\label{eq:5dds20}} where the five dimensional emblackening factor is given by \eq{f(r) = 1 - \frac{\m}{r^2}+\frac{r^2}{\rads^2},\label{eq:5demblackening}} where the mass parameter $\m$ is related to the mass $M$ of the black hole by \eq{\m = \frac{8GM}{3\pi}.} The event horizon is located at \eq{r_H =  \sqrt{\frac{\rads^2}{2}\left(\sqrt{1+\frac{4\m}{\rads^2}}-1\right)}\label{eq:rH}.}

\subsection{CV for Einstein Gravity}
\begin{figure}
\begin{centering}
\includegraphics[scale=1]{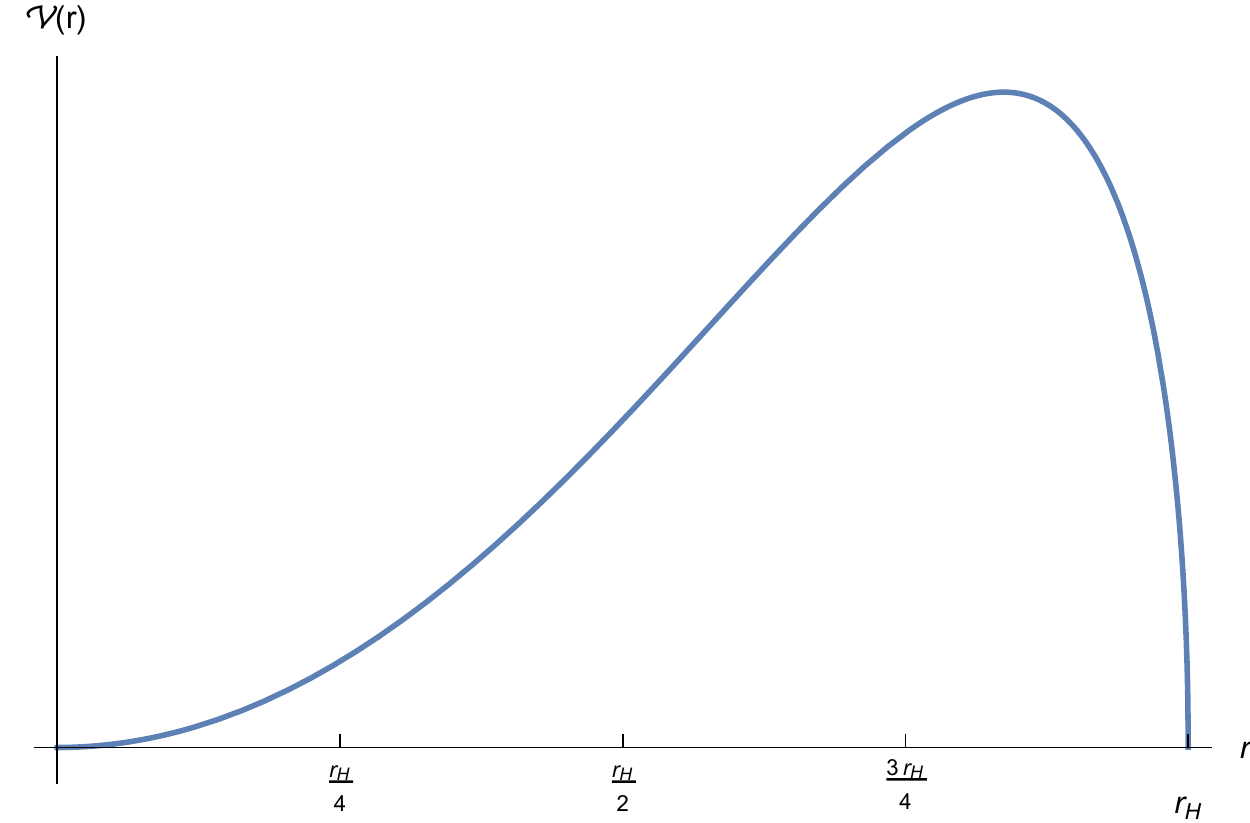}
\caption{The ``volume functional" $\calV(r)$ for the AdS-Schwarzschild metric in Eq. \ref{eq:5dds20}.}
\label{fig:cvGR}
\end{centering}
\end{figure}

We will begin with the CV conjecture \cite{Susskind:2014rva,Stanford:2014jda,Susskind:2014moa}. Our starting point is the metric in Eqs. \ref{eq:5dds20} and \ref{eq:5demblackening}, from which we can straightforwardly compute $\calV(r)$, which is plotted in Figure \ref{fig:cvGR}, and extremize to find that\eq{r_* = \frac{1}{2\sqrt{2}}\sqrt{\rads\left(-3\rads+\sqrt{32\m+9\rads^2}\right)}.\label{eq:r0}} Plugging in to Eq. \ref{eq:VdotDef}, we have that \eq{\vdot = \frac{\Om_3}{64} \left(\sqrt{9 \rads^4+32 \mu  \rads^2}-3 \rads^2\right)^{3/2} \sqrt{\frac{\sqrt{9 \rads^4+32 \mu  \rads^2}}{\rads^2}+1}} and therefore that \eq{\cdot_V = \frac{\Om_3}{64G\rads} \left(\sqrt{9 \rads^4+32 \mu  \rads^2}-3 \rads^2\right)^{3/2} \sqrt{\frac{\sqrt{9 \rads^4+32 \mu  \rads^2}}{\rads^2}+1}.\label{eq:cvGR}}  This is exact, but fairly messy; to compare to the simple form of the answer in \cite{Stanford:2014jda}, it is convenient to take the high temperature limit $\m\gg1$, which corresponds to dropping the one in Eq. \ref{eq:5demblackening}. In this case one finds \eq{r_* = \sqrt{\rads\sqrt{\frac{\m}{2}}}} and therefore that \eq{\calV\prs = \frac{\m\rads}{2}.} Putting it all together, we have that \eq{\vdot = \frac{8\piG M\rads}{3},} which exactly matches \cite{Stanford:2014jda}. However, in this paper we will not work in the high temperature limit, and will instead work with the full metric in Eq. \ref{eq:5demblackening}. 

\subsection{CA for Einstein Gravity}
We will now move on to the CA conjecture \cite{Brown:2015bva,Brown:2015lvg}. Our goal here is to use Eq. \ref{eq:SWdW} to compute the action of the $\de$WdW patch shown in Figure \ref{fig:deWdW}. For general relativity, the bulk action in Eq. \ref{eq:SWdW} is the usual Einstein Hilbert action action, \eq{S_{\text{GR, bulk}} = \frac{1}{16\piG}\int d^5x\sqrt{-g}\left[R-2\L\right],} and the boundary action $S_{\text{bdy}}$ is the Gibbons-Hawking-York (GHY) boundary term, which for spacelike hypersurfaces is given by \eq{S_{\text{bdy}} = \frac{1}{8\piG}\int d^{d-1}x\sqrt{-h}K,} where $h_{ab}$ is the induced metric on the boundary and $K$ is the trace of the extrinsic curvature tensor.

The bulk part is fairly simple; we have \eq{R = -\frac{20}{\rads^2}} so that \eq{S_{\text{GR, bulk}} = \frac{1}{16\piG}\int d^5x\sqrt{-g}\left[R-2\L\right] = -\frac{\de{t}\Om_3}{16\piG}\int_\eps^\rh dr\ r^3\left[\frac{8}{\rads^2}\right] = -\frac{\de{t}\Om_3}{8\rads^2\piG}\left[r_H^4-\eps^4\right].} Now we need to evaluate the boundary terms. A standard calculation shows that, for a constant-$r$ hypersurface, we have \eq{K = \frac{\sqrt{f(r)}}{2}\left(\frac{f'(r)}{f(r)} + \frac{6}{r}\right),} from which we can easily see that {\small \eq{S_{\text{GR, bdy}} = \frac{\de{t}\Om_3}{8\piG}\sqrt{f(r)}r^3\left[\frac{\sqrt{f(r)}}{2}\left(\frac{f'(r)}{f(r)} + \frac{6}{r}\right)\right] = \frac{\de{t}\Om_3}{16\piG}r^3\left(f'(r) + \frac{6f(r)}{r}\right) = \frac{\de{t}\Om_3}{16\piG}\left(6r^2+\frac{8r^4}{\rads^2}-4\m\right).}} Putting it all together, and plugging in Eq. \ref{eq:rH}, we have that \eq{S_{\de{WdW}} = 2M\de{t}} and therefore that \eq{\cdot_A = \frac{S_{\de{WdW}}}{\pi\de{t}} = \frac{2M}{\pi},} exactly matching both the famous result of \cite{Brown:2015bva,Brown:2015lvg} and the Lloyd bound in Eq. \ref{eq:lloyd}. 

\addcontentsline{toc}{section}{References}
\bibliographystyle{JHEP}
\bibliography{apBib}

\end{document}